\documentclass[sigconf]{acmart}

\AtBeginDocument{%
  \providecommand\BibTeX{{%
    Bib\TeX}}}

\def\BibTeX{{\rm B\kern-.05em{\sc i\kern-.025em b}\kern-.08emT\kern-.1667em\lower.7ex\hbox{E}\kern-.125emX}}

\newcommand{\labelnoisenerf}{\textsf{NoisyLabelNeRF}\xspace}
\newcommand{\splitnerf}{\textsf{SplitNeRF}\xspace}
\newcommand{\securesplitnerf}{\textsf{$S^2$NeRF}\xspace}

\newcommand{\reconstructattack}{\textsf{Surrogate Model Attack}\xspace}
\newcommand{\finetuneattack}{\textsf{Scene-aided Surrogate Model Attack}\xspace}

\usepackage{nicefrac} 
\usepackage{siunitx}
\usepackage{array,framed}
\usepackage{multirow}
\usepackage{booktabs}
\usepackage{
  color,
  float,
  epsfig,
  wrapfig,
  graphics,
  graphicx,
  subcaption
}

\usepackage[linesnumbered,boxed,ruled,commentsnumbered]{algorithm2e}
\usepackage{algorithmic}
\usepackage{amsmath}
\usepackage{comment}
\usepackage{amssymb}
\usepackage{graphicx}
\usepackage{colortbl} 
\usepackage{xcolor}
\usepackage{array}   
\usepackage{xspace}
\usepackage{balance}
\usepackage{booktabs}
\usepackage{enumitem}
\usepackage{arydshln}
\usepackage{makecell}
\setlist[itemize]{leftmargin=*}

\usepackage{
  tikz,
  pgfplots,
  pgfplotstable
}
\usepackage{hyperref}
\newcommand{\mypara}[1]{\smallskip\noindent\textbf{#1.}\xspace}

\hypersetup{
  colorlinks,
  linkcolor={blue!70!green},
  citecolor={green!70!blue},
  urlcolor={orange!70!red}
}

\usetikzlibrary{
  shapes.geometric,
  arrows,
  external,
  pgfplots.groupplots,
  matrix
}

\pgfplotsset{compat=1.9}

\usepackage{float}
\usepackage{mathtools,}

\DeclareMathAlphabet{\mathcal}{OMS}{cmsy}{m}{n}

\DeclareGraphicsExtensions{%
    .png,.PNG,%
    .pdf,.PDF,%
    .jpg,.mps,.jpeg,.jbig2,.jb2,.JPG,.JPEG,.JBIG2,.JB2}

\setlist[itemize]{leftmargin=*}    
\copyrightyear{2024}
\acmYear{2024}
\setcopyright{rightsretained}
\acmConference[CCS '24]{Proceedings of the 2024 ACM SIGSAC Conference on Computer and Communications Security}{October 14--18, 2024}{Salt Lake City, UT, USA}
\acmBooktitle{Proceedings of the 2024 ACM SIGSAC Conference on Computer and Communications Security (CCS '24), October 14--18, 2024, Salt Lake City, UT, USA}
\acmDOI{10.1145/3658644.3690185}
\acmISBN{979-8-4007-0636-3/24/10}

\begin{document}

\title{\securesplitnerf: Privacy-preserving Training Framework for NeRF}

\author{Bokang Zhang}
\authornote{Bokang and Yanglin are co-first authors.}
\affiliation{%
  \institution{The Chinese University of Hong Kong, Shenzhen}
  \city{}
  \country{}}
\email{bokangzhang@link.cuhk.edu.cn}

\author{Yanglin Zhang}
\authornotemark[1]
\affiliation{%
  \institution{The Chinese University of Hong Kong, Shenzhen}
  \city{}
  \country{}}
\email{yanglinzhang@link.cuhk.edu.cn}

\author{Zhikun Zhang}
\authornote{Corresponding authors.}
\affiliation{%
  \institution{Zhejiang University}
  \city{}
  \country{}}
\email{zhikun@zju.edu.cn}

\author{Jinglan Yang}
\affiliation{%
  \institution{The Chinese University of Hong Kong, Shenzhen}
  \city{}
  \country{}}
\email{jinglanyang@link.cuhk.edu.cn}

\author{Lingying Huang}
\affiliation{%
  \institution{Nanyang Technological University}
  \city{}
  \country{}}
\email{lingying.huang@ntu.edu.sg}

\author{Junfeng Wu}
\authornotemark[2]
\affiliation{%
  \institution{The Chinese University of Hong Kong, Shenzhen}
  \city{}
  \country{}}
\email{junfengwu@cuhk.edu.cn}

\renewcommand{\shortauthors}{Zhang et al.}

\begin{abstract}
\textit{Neural Radiance Fields} (NeRF) have revolutionized 3D computer vision and graphics, facilitating novel view synthesis and influencing sectors like extended reality and e-commerce. 
However, NeRF's dependence on extensive data collection, including sensitive scene image data, introduces significant privacy risks when users upload this data for model training.
To address this concern, we first propose a strawman solution: \splitnerf, a training framework that incorporates \textit{split learning} (SL) techniques to enable privacy-preserving collaborative model training between clients and servers without sharing local data. 
Despite its benefits, we identify vulnerabilities in \splitnerf by developing two attack methods, \reconstructattack and \finetuneattack, which exploit the shared gradient data and few leaked scene images to reconstruct private scene information. 
To counter these threats, we introduce \securesplitnerf, secure \splitnerf that integrates effective defense mechanisms. By introducing decaying noise related to the gradient norm into the shared gradient information, \securesplitnerf preserves privacy while maintaining a high utility of the NeRF model.
Our extensive evaluations across multiple datasets demonstrate the effectiveness of \securesplitnerf against privacy breaches, confirming its viability for secure NeRF training in sensitive applications.\footnote{The implementation can be found at \url{https://github.com/lucky9-cyou/S2-NeRF}.}
\end{abstract}

\ccsdesc[500]{Security and privacy}
\ccsdesc[500]{Computing methodologies~Machine learning}

\keywords{Privacy-preserving Machine Learning; Split Learning; Neural Radiance Fields (NeRFs)}

\maketitle

\section{Introduction}
\textit{Neural radiance fields} (NeRF)~\cite{mildenhall2020nerf} represent a transformative development in the domain of novel view synthesis, catalyzing significant advancements in 3D computer vision and graphics. 
The impact of NeRF is wide-ranging, influencing various sectors such as extended reality, 3D generative AI, e-commerce, and robotics. 
NeRF's training procedure necessitates the collection of camera parameters, including position and orientation, alongside images captured from varied camera positions. 
This technology is increasingly incorporated into consumer devices, such as extended reality glasses, to model users' environments accurately.
Presently, to facilitate NeRF model training, users are often required to upload images of their surroundings along with camera parameters to a cloud server managed by the service provider~\cite{li2024disorf}. 
This process, while standard, allows the company unrestricted access to the uploaded images. 
Moreover, the trained NeRF models enable these companies to reconstruct user environments in extensive detail, posing significant privacy risks. 
Such exposure of personal data underscores a critical need for enhanced privacy safeguards in NeRF applications.

The privacy concerns associated with NeRF can be effectively addressed using \textit{split learning} (SL), a methodology that suits the privacy requirements essential for training neural networks collaboratively between two parties~\cite{gupta2018distributed, vepakomma2018split, thapa2022splitfed}. 
This approach is particularly pivotal in \textit{vertical federated learning} (VFL), where the data is vertically partitioned into two parties~\cite{feng2020multi}. 
Within VFL, there are primarily two architectures: one without model splitting~\cite{fu2022label} and another that incorporates model splitting~\cite{vepakomma2018split}. 
SL is categorized under VFL which includes model splitting.
Within the framework of split learning, two distinct parties, the “user party” possessing the data inputs and the “label party” holding the labels, can cooperatively train a model comprising two sub-models—“user model” and “label model”. 
This setup ensures that neither party has to reveal their respective data inputs or labels to the other. The only information exchanged between the two parties involves the cut layer embeddings, used for forward computation, and the gradients, necessary for the backward updates, thereby safeguarding the privacy of both the data and labels~\cite{vepakomma2018split}.

Recent research has uncovered several privacy attacks in SL settings, demonstrating that sensitive labels can be accurately inferred by adversarial parties across classification and regression tasks~\cite{erdougan2022unsplit, li2021label, fu2022label, xie2023label, pasquini2021unleashing}.
These studies highlight a significant risk in SL environments: the potential for an adversarial user party to reconstruct sensitive label information using access to partial models and shared gradient data, underscoring the need for enhanced security measures in SL systems.

To address such privacy concerns, privacy protection is primarily achieved through two categories of methods. 
The first involves the use of sophisticated cryptographic protocols, such as \textit{secure multiparty computation}~\cite{wagh2020falcon, mohassel2018aby3} and \textit{two-party computation}~\cite{mohassel2017secureml, patra2021aby2, xie2021generalized}. 
These approaches provide strong security guarantees by ensuring that the computation on private inputs is performed without revealing them to any party.
The second category employs perturbation-based methods that introduce randomness into the data-sharing process to protect privacy. 
This includes obfuscating the shared gradient information among parties or directly perturbing the information that needs protection~\cite{abadi2016deep, erlingsson2019amplification, ghazi2021deep, sun2023dpauc, yang2022differentially, xie2022differentially, qiu2023defending}.

\mypara{Our Contributions}
In this paper, we introduce \splitnerf, a NeRF training framework adapted to the SL methodology.
In \splitnerf, the client and server collaboratively train the NeRF model without exchanging local client data, addressing the inherent privacy concerns of traditional NeRF training approaches.
To empirically assess the privacy risks of the \splitnerf framework, we propose and implement two NeRF attack methods: \reconstructattack and \finetuneattack. 
We implement the \reconstructattack method by integrating the model reconstruction attack~\cite{fu2022label} with the gradient matching technique~\cite{xie2023label}. 
We employ constructed dummy image data to enhance the surrogate model's ability to approximate the client's model.
To explore the effectiveness of \reconstructattack, we conduct extensive experiments utilizing various learning rate strategies across three popular datasets. 
These experiments demonstrate its capability to accurately restore the geometric outlines of the original scenes, albeit in black-and-white format. 
Moreover, the \finetuneattack addresses scenarios where an attacker has access to a limited number of actual scene images, a common occurrence given the proliferation of social networks. 
\finetuneattack leverages leaked image data to determine the corresponding camera poses, using this information to fine-tune the attack model trained by \reconstructattack.
This method effectively enhances the reconstructed black-and-white images to restore their color, achieving a Peak Signal Noise Ratio(PSNR) of $27.62$ and a Learned Perceptual Image Patch Similarity(LPIPS)~\cite{zhang2018unreasonable} of $0.19$, demonstrating significant restoration quality even with just one leaked image.

To remedy these vulnerabilities, we develop \securesplitnerf, secure \splitnerf that incorporates robust defense mechanisms.
\securesplitnerf distorts the gradient information transmitted from the client to the server. 
By perturbing the gradients, it safeguards the client's scene privacy by disrupting the process of \reconstructattack and \finetuneattack. 
To ensure \securesplitnerf's applicability across various scenarios, \securesplitnerf adjusts the noise scale based on the norm of the transmitted gradient and employs a decay strategy for the noise. 
The decay strategy is designed to balance privacy protection with the utility of the NeRF model during training.
We have conducted comprehensive experiments to assess the effectiveness of \securesplitnerf against multiple attack strategies across three datasets. 
These experiments demonstrate that with appropriate configuration, \securesplitnerf can effectively resist attacks while maintaining high model utility across different dataset environments. 
This balance showcases the potential of \securesplitnerf to serve as a viable solution for secure NeRF training in diverse applications.

\begin{itemize}
    \item \mypara{(C1) \splitnerf}
    To our knowledge, \splitnerf is the first approach that tackles privacy issues in NeRF training by employing SL techniques. 
    Within the \splitnerf framework, the client and server collaboratively train the NeRF model without necessitating the transfer of local private data to the server. 

    \item \mypara{(C2) \reconstructattack and \finetuneattack} 
    We develop two attack strategies, \reconstructattack and \finetuneattack against the \splitnerf framework. 
    Comprehensive experiments on three indoor NeRF datasets demonstrate that these attacks can effectively exploit vulnerabilities in the vanilla \splitnerf setup, revealing significant privacy risks. 

    \item \mypara{(C3) \securesplitnerf} 
    To remedy the vulnerabilities uncovered, we propose the secure \splitnerf framework, which we call \securesplitnerf.
    Our extensive experiments have shown that \securesplitnerf effectively prevents attackers from attempting to recreate any part of the scene while maintaining the high utility of the NeRF model.
\end{itemize}


\section{Preliminaries} \label{sec:preliminary}

\subsection{NeRF in 3D Reconstruction} \label{subsec:nerf-pre}
NeRF~\cite{mildenhall2020nerf} uses \textit{multilayer perception} (MLP) $\Theta_\sigma$ and $\Theta_c$ to map the 3D location $\mathbf{x} \in \mathbb{R}^3$ and viewing direction $\mathbf{d} \in \mathbb{R}^2$ to a RGB color $\mathbf{c} \in \mathbb{R}^3$ and a density value $\sigma \in \mathbb{R}^{+}$:
$$
\begin{gathered}
{[\sigma, \mathbf{z}]=\Theta_\sigma\left(\gamma_{\mathbf{x}}(\mathbf{x})\right),} \\
\mathbf{c}=\Theta_c\left(\mathbf{z}, \gamma_{\mathbf{d}}(\mathbf{d})\right),
\end{gathered}
$$
where $\gamma_{\mathbf{x}}$ and $\gamma_{\mathbf{d}}$ are high-frequency positional encoding for location and viewing direction, respectively, which are designed to make MLP better suited for modeling functions in low dimensions.
It's able to overcome the spectral bias inherent in the NeRF model~\cite{tancik2020fourier}.
The intermediate variable $\mathbf{z}$ is a feature output by the first MLP $\Theta_\sigma$.
To better represent the space and improve training efficiency, part of the MLPs $\Theta_\sigma$ can be replaced with a small neural network~\cite{mueller2022instant}, which is augmented by a multi-resolution hash table of trainable feature vectors optimized through stochastic gradient descent.

For rendering a 2D image from the radiance fields $\Theta_\sigma$ and $\Theta_c$, a numerical quadrature is used to approximate the volumetric projection integral. 
Formally, $N_p$ points are sampled along a camera ray $r$ with color and geometry values $\left\{\left(\mathbf{c}_r^i, \sigma_r^i\right)\right\}_{i=1}^{N_{p}}$. 
The RGB color value $\hat{\mathbf{C}}(r)$ is obtained using alpha composition~\cite{porter1984compositing}
$$
\hat{\mathbf{C}}(r)=\sum_{i=1}^{N_p} T_r^i\left(1-\exp \left(-\sigma_r^i \delta_r^i\right)\right) \mathbf{c}_r^i,
$$
where $T_r^i=\prod_{j=1}^{i-1}\left(\exp \left(-\sigma_r^j \delta_r^j\right)\right)$, and $\delta_r^i$ is the distance between adjacent sample points. 
The MLPs $\Theta_\sigma$ and $\Theta_c$ are optimized by minimizing the reconstruction loss between observations $\mathbf{C}$ and predictions $\hat{\mathbf{C}}$ as
\begin{equation} \label{eq:loss_reconstruction}
    L_{recon} = \frac{1}{N_r} \sum_{m=1}^{N_r}\left\|\hat{\mathbf{C}}\left(r_m\right)-\mathbf{C}\left(r_m\right)\right\|_2^2,
\end{equation}
where $N_r$ is the number of sampled pixels.
Given $\Theta_\sigma$ and $\Theta_c$, novel views are synthesized by invoking volume rendering for each ray.

\begin{table}[!tbp]
	\centering 
	\caption{\textbf{Summary of the notations.}} 
        \setlength{\tabcolsep}{1mm}
	\begin{tabular}{c|l}  
        \toprule
		\textbf{Notations} & \textbf{Descriptions} \\ 
        \midrule
		$\mathbf{x,d}$ & 3D location and viewing direction \\
        $\mathbf{c}, \sigma$ & RGB Color and density of a 3D point  \\
        $\Theta_\sigma, \Theta_c$ & MLPs that output $\sigma$ and $\mathbf{c}$ \\
        $\gamma_{\mathbf{x}}$, $\gamma_{\mathbf{d}}$  & High-frequency positional encoding from $\mathbf{x,d}$ \\
        $r$ & A camera ray from the camera position to a pixel\\
        $N_p$ & Number of sampled points in a ray \\     
        $\left\{\left(\mathbf{x}_r^i, \mathbf{d}_r^i\right)\right\}_{i=1}^{N_{p}}$ & 
        The set of 3D points sampled along a camera ray $r$ \\ 
        $\left\{\left(\mathbf{c}_r^i, \sigma_r^i\right)\right\}_{i=1}^{N_{p}}$
        & {The set of 3D points' RGB color and density} \\
        $N_r$ & Number of sampled pixels \\  
        $\hat{\mathbf{C}}(r), \mathbf{C}(r)$ & RGB predictions and ground truth of ray $r$ \\
        \bottomrule 
	\end{tabular}
    \label{table:notations}  
\end{table}

\subsection{Two-party Split Learning} \label{subsec:splitlearn-pre}

\begin{figure*}[!tbp]
\centering
\includegraphics[width=170mm]{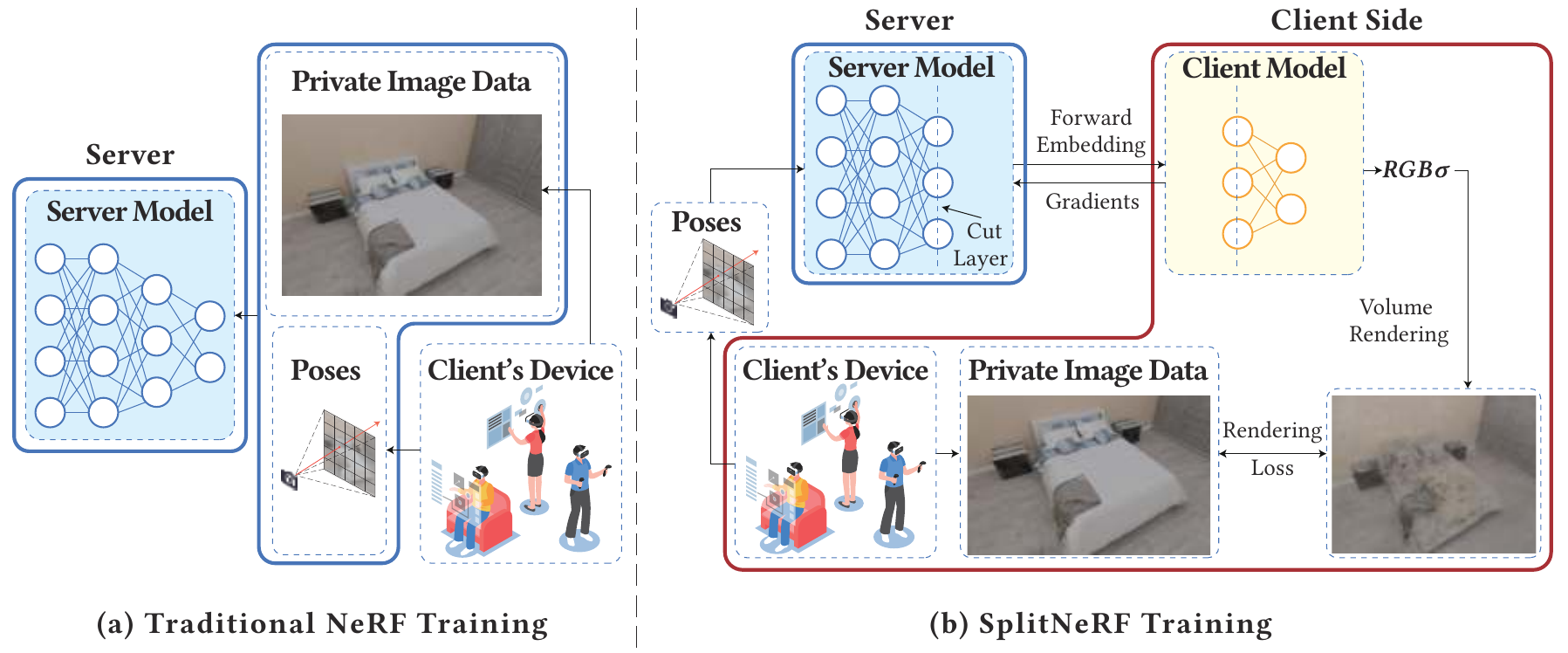}
\caption{
\textbf{Overview of \splitnerf framework.
Traditional NeRF training requires the uploading of all training data, including scene images and corresponding camera poses, to a central server, presenting a substantial risk to privacy. 
We propose a split learning-based NeRF training framework, named \splitnerf.
The entire NeRF training model is divided into two parts, the server model and the client model.
Specifically, clients are required only to send camera poses to the server, while keeping private image data local.
During training, the client transmits a series of poses to the server, which then calculates and sends back the embeddings of these poses. 
The client proceeds to compute the colors and densities associated with these poses, followed by rendering and loss function calculations. 
The client then sends the gradients back to the server, thus completing the backpropagation process.
}}
\label{fig:splitnerf}
\end{figure*}

Split learning~\cite{gupta2018distributed, vepakomma2018split} has been proposed to enable two parties, a user party and a label party, to collaboratively train a composite model with the vertically partitioned data (usually in federated learning~\cite{yang2019federated}). 
In particular, the user party holds the feature of the training data while the label party holds the label.
By the data location, the composite model is split into the user model and label model held by the user and label party respectively.
More formally, for user party, denote the user's input data set $X=\left\{x_i, i \in[1, \ldots, |D|]\right\}$, and the user model $\mathcal{M}_{user}$; for the label party, denote the label party's label $Y=\left\{y_i, i \in[1, \ldots, |D|]\right\}$ and label model $\mathcal{M}_{label}$.
$D=\left\{\left(x_i, y_i\right)\right\}$ is the whole training dataset to use for training the composite model $\mathcal{M}_{user} \circ \mathcal{M}_{label}$.
For example, in the practical advertisement conversion prediction, $x_i$ can be a feature vector about a user and $y_i$ indicates whether the user clicks on the advertisement.
The final layer of the user model $\mathcal{M}_{user}$, where the composite model is split, is called the cut layer.

Split learning follows the conventional training process,  consisting of two phases: 1) forward propagation and 2) backpropagation.

\mypara{Forward Propagation} 
The user party computes the embedding of the cut layer $\mathcal{E}_i=M_{user}\left(x_i\right)$. 
This embedding is used as input to the label model held in the label party to compute the prediction score of the whole composite model:
$$
y_i^{predict}=\mathcal{M}_{label}\left(\mathcal{M}_{ user }\left(x_i\right)\right)=\mathcal{M}_{label }\left(\mathcal{E}_i\right).
$$
Then we evaluate the loss to capture the disagreement between the predictions and the ground truth: $L\left(y_i^{predict}, y_i\right)$.

\mypara{Backpropagation} 
The label party first updates its label model's parameters by computing the gradient of loss $L$ concerning the label model itself $\mathcal{M}_{label}$. 
To further help the user party to compute the updates for user model $\mathcal{M}_{user}$, the label party needs to compute the gradient for the embedding $\mathcal{E}_i$ at the cut layer by a chain rule, $\nabla_{\mathcal{E}_i} L$ (denoted as $g$ for short), where
$$ 
g \triangleq \nabla_{\mathcal{E}_i} L=\frac{\partial L\left(y_i^{predict}, y_i\right)}{\partial y_i^{predict}} \cdot \frac{\partial y_i^{predict}}{\partial \mathcal{E}_i}.
$$
Then the label party sends this gradient to the user party.
Finally, the user party continues to compute the gradient for the user model w.r.t. $\mathcal{M}_{user }$ 's parameters:
$$
\nabla_{\mathcal{M}_{user}} L=\nabla_{\mathcal{E}_i} L \cdot \frac{\partial \mathcal{E}_i}{\partial \mathcal{M}_{user}}.
$$

As for the inference phase, the user party first computes the cut layer embedding $\mathcal{E}_i=\mathcal{M}_{user }\left(x_i\right)$ and sends it to the label party, which computes the final prediction.

\subsection{Notations} \label{subsec:notations}
Frequently used notations in SL and NeRF are summarised in \autoref{table:notations}.

\section{The Strawman Solution: \splitnerf}
\label{sec:splitnerf} 
Cloud computing facilitates NeRF training between an edge device and a remote server.
The edge device, which is resource-constrained and lacks support for powerful GPUs and extensive memory capacities, acts as the client of the training service provided by the server. 
In standard training configurations, the client needs to upload image data to the server, posing potential privacy risks.
To address the privacy leakage concerns, we first propose the strawman solution: \splitnerf, a split learning framework tailored for NeRF training in 3D reconstruction.
Compared to traditional centralized NeRF training, \splitnerf keeps the labels locally and sends the features to the server. 
In the context of NeRF training, the features (camera poses) are not considered private information.

The \splitnerf framework is illustrated in \autoref{fig:splitnerf}.
It assigns a portion of the NeRF model to the cloud server, therefore alleviating the computational burden on the client. 
During the \splitnerf training process, the server does not directly gain access to the client's private image dataset but solely contributes computational resources. 
All sensitive image data is strictly retained on the client side.
Within this framework, the NeRF model is partitioned into two components: the server model $\mathcal{M}_{server}$ and the client model $\mathcal{M}_{client}$, by which the MLP layer responsible for generating density information must be retained on the client side. 
This setup enables the client to retain privacy-sensitive data locally, thus bolstering data privacy.

Following the standard NeRF training process and the virtue of split learning, \splitnerf consists of two phases: 1) forward propagation and 2) backpropagation.

\mypara{Forward Propagation} 
Initially, $N_p$ points $\left\{\left(\mathbf{x}_r^i, \mathbf{d}_r^i\right)\right\}_{i=1}^{N_{p}}$ are sampled along a camera ray $r$ on the client side. 
Subsequently, these points are transmitted to the server.
The server then calculates the cut layer embeddings for these $N_p$ points, denoted as $\left\{\mathcal{E}_r^i \right\}_{i=1}^{N_{p}}$. These embeddings are sent back to the client.
Upon receiving the embeddings, the client uses them as inputs to its model to compute the final output labels, which include predicted density and color, represented as $\left\{\left(\mathbf{c}_r^i, \sigma_r^i\right)\right\}_{i=1}^{N_{p}}$.
The client then calculates the predicted RGB color $\hat{\mathbf{C}}(r)$. 
The reconstruction loss between the predictions and the ground truth is then calculated using the local image data held by the client.

\mypara{Backpropagation} 
After calculating the reconstruction loss, the client proceeds to update its model $\mathcal{M}_{client}$. 
To facilitate the server in completing updates to its model parameters $\mathcal{M}_{server}$, the client computes the shared gradients $g_r^i = \frac{\partial Loss}{\partial \mathcal{E}_r^i}$ for the cut layer embeddings using a chain rule. 
With these shared gradients, the server is then able to complete the backpropagation process for its network $\mathcal{M}_{server}$.

In the \splitnerf inference, the client is tasked with rendering an image based on the 3D position and orientation of a camera. 
If the client's storage capacity permits, the server may transfer the entire set of server model parameters to the client,  excluding itself from the inference process.
Alternatively, if the client lacks sufficient local computing resources, the inference can be conducted interactively. 
In this scenario, the client begins by sampling the points required for rendering, based on the camera's position and orientation, and sends these points to the server. 
Then,  similar to the forward propagation process, the client obtains the predicted pixels and proceeds to render the image.

\mypara{Privacy Analysis} 
\splitnerf enhances privacy compared to traditional NeRF training methods that require uploading local datasets to a cloud server. 
By keeping the image dataset local, \splitnerf minimizes data exposure. 
Nevertheless, the server retains access to portions of the NeRF network parameters and receives gradient information during training. This raises an important question: \textit{could the server exploit this information to compromise local data privacy?} 
Next, we evaluate this privacy concern by devising attack methods against \splitnerf to assess its vulnerabilities.


\begin{figure*}[!tbp]
\centering
\includegraphics[width=170mm]{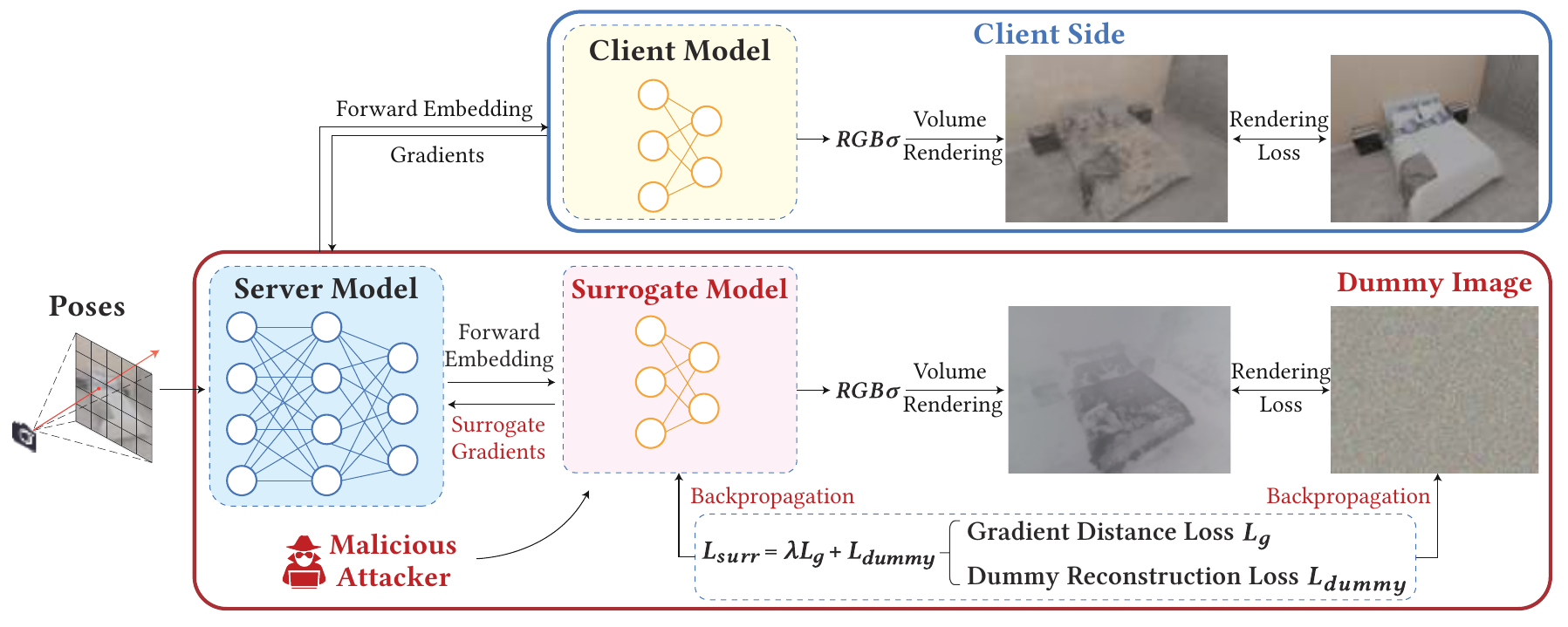}
\caption{\textbf{Overview of \reconstructattack against \splitnerf.
The attacker(curious server) attempts to mimic the representation capabilities of the client model to generate a 3D description of the client's private scenario.
To achieve this, the attacker sets up a surrogate model along with dummy image data.
The attacker utilizes backpropagation to optimize both the surrogate model and dummy image with the surrogate model loss $L_{surr}$, which includes gradient distance loss $L_g$ and dummy reconstruction loss $L_{dummy}$.}
}
\label{fig:nerf-attack}
\end{figure*}

\section{Privacy Risk Assessment of \splitnerf}
\label{sec:attack}

In this section, we comprehensively assess the privacy risks of \splitnerf by designing two attacks to restore the client's NeRF model to infer the client's private information.
First, we introduce \reconstructattack against \splitnerf, and then we introduce \finetuneattack provided the attacker is additionally capable of getting access to a few leaked scene image data.

\subsection{\reconstructattack} \label{subsec:ModelCompletionAttack}

\indent
\subsubsection{Threat Model} \label{subsubsec:threatmodel}
We define the threat model for our attack in aspects of security research, including the attack setting and goal, and the attacker's capability and knowledge. 

\mypara{Attack Setting \& Goal}
Similar to numerous other studies in SL~\cite{li2021label, fu2022label}, we employ the honest-but-curious model assumption. In this model, the attacker (server) adheres to the standard training protocol while attempting to deduce the private information and neural networks on the client side.
In NeRF applications, the user's network parameters are often as sensitive as the user's stored image data, unlike in traditional deep-learning tasks, as the NeRF neural network serves as a comprehensive neural representation of the client's scene.
If an attacker gains access to the client model, it could render views of the client's environment from any 3D poses by integrating the client model with the server model. 
Consequently, the attacker's main goal is to create a surrogate model that closely matches the client model.

\mypara{Attackers’ Capabilities \& Knowledge}
\reconstructattack operates under a black-box setting, where the attacker is unaware of both the client model parameters and the client model structure.
In this scenario, the attacker has the flexibility to tailor its surrogate model, for example, by increasing its depth or width, to facilitate the attack. 
Unlike other SL scenarios~\cite{fu2022label, xie2023label}, obtaining even a small portion of complete training data is challenging in NeRF training due to the difficulty in accurately matching images to their corresponding poses. 
This is explored further in \autoref{subsec:attack_advanced}.
Consequently, the attacker must construct the attack based solely on the server model parameters and the gradient information exchanged during training. 
The attacker has the discretion to determine the training approach for the surrogate model based on the available information.

\subsubsection{Attack Methodology} \label{subsubsec:attackmethod}
Our proposed \reconstructattack aims to approximate the representation capabilities of the client model through a learning-based method.
Initially, we construct a surrogate model designed to approximate the client model and dummy image data to approximate the client's image data. 
We then establish a loss function incorporating available information, such as gradients, and integrate additional machine learning objectives, such as model performance, as forms of regularization.
Subsequently, we iteratively backpropagate the surrogate model and the dummy image data.
This approach is detailed in \autoref{fig:nerf-attack}, illustrating our attack framework.

The surrogate model and the corresponding dummy image data for a given ray $r$ are denoted as $M_{surr}$ and ${\mathbf{C}}(r)_{dummy}$, respectively.
We first initialize $M_{surr}$ and ${\mathbf{C}}(r)_{dummy}$ and then calculate the predictions made by the surrogate model and the gradients at the cut layer (denoted as $g_{surr}$) following the \splitnerf process.
Next, we introduce the loss function for optimizing the surrogate model and dummy image data.

\begin{algorithm}[!tbp]
	\caption{\reconstructattack against \splitnerf}
	\label{alg:SplitNeRFAttack}
	\textbf{Input:} Training data(rays) $\left\{r_m, m \in[1, \ldots, |N_r|]\right\}$, Server model $\mathcal{M}_{server}$, Number of training epochs $T$, surrogate model learning rate $\eta_t$ \\
    \textbf{Output:} Inferred surrogate model $\mathcal{M}_{surr}$\\
    \textbf{Procedures:}
    \begin{algorithmic}[1]
    \STATE \text{/* Initialize dummy image data. */}
    \STATE ${\mathbf{C}}_{dummy}^{(0)} \leftarrow [\mathcal{N}(0,1)]^{|N_r|}.$
    \STATE Initialize the surrogate model $\mathcal{M}_{surr}^{(0)}$.
    \FOR{$t = 0$ to $T$} 
        \STATE \text{Compute $L_g$ and $L_{dummy}$ in \eqref{eq:loss_gradient} and \eqref{eq:loss_dummy}}
        \STATE $L_{surr } \leftarrow \lambda L_g + L_{dummy}$;
        \STATE \text{/* Optimize the surrogate model and dummy image data;*/}
    \STATE $\mathcal{M}_{surr}^{(t+1)} \leftarrow  \text{AdamUpdate}(\mathcal{M}_{surr}^{(t)}$);
    \STATE ${\mathbf{C}}_{dummy}^{(t+1)} \leftarrow \text{AdamUpdate}({\mathbf{C}}_{dummy}^{(t)}$);
    \ENDFOR
    \end{algorithmic}
\textbf{Return:} $\mathcal{M}_{surr}^{(T+1)}$ as surrogate model
\end{algorithm}

\mypara{Gradient Distance Loss}
To reconstruct the private client's network, we primarily utilize the shared gradient used in backpropagation. 
Specifically, the gradient $g_{surr}$ produced by the surrogate model should closely approximate the original gradient $g$ generated by the client model. 
To measure the closeness between these two gradients, we introduce the gradient distance loss, which quantifies the discrepancy between $g$ and $g_{surr}$:
\begin{equation} \label{eq:loss_gradient}
    L_g=\mathcal{D}\left(g, g_{surr}\right),
\end{equation}
where $\mathcal{D}(\cdot)$ denotes a distance function, e.g., the $\ell_2$ norm function. 

\mypara{Dummy Reconstruction Loss}
Considering that our learning objective involves high-dimensional spaces~\cite{xie2023label}, it could be very likely to get different surrogate model parameters even if the gradient distance loss is close to $0$.
To constrain our learning space and guide the surrogate model to behave like the client model, we incorporate a dummy reconstruction loss as regularization.
By incorporating this loss term, the surrogate model is compelled to function like the original NeRF model, particularly in scene reconstruction performance.
Accordingly, we define the dummy reconstruction loss to ensure that the predictions of the ray $r$ by the surrogate model $M_{surr}$ closely match the dummy image data ${\mathbf{C}}(r)_{dummy}$ once the surrogate model has converged.

Specifically, for a given ray $r$, the surrogate model $M_{surr}$ first predicts the density and colors of the 3D sampled points along $r$ and then computes the RGB color $\hat{\mathbf{C}}(r)_{surr}$ at the pixel $r$.
We then define the  reconstruction loss as follows:
\begin{equation} \label{eq:loss_dummy}
L_{dummy} = \frac{1}{N_r} \sum_{m=1}^{N_r}\left\| \hat{\mathbf{C}}(r_m)_{surr} - {\mathbf{C}}(r_m)_{dummy} \right\|_2^2,
\end{equation}
where $N_r$ represents the number of sampled pixels.

Additionally, we incorporate weighting parameters $\lambda$ to balance the functionality of the components of the learning loss.
In a nutshell, the learning loss for the surrogate model is given by
\begin{equation} \label{eq:loss_surrogate}
    L_{surr} = \lambda L_g + L_{dummy}.
\end{equation}
To minimize the above loss function over the surrogate model, we employ a gradient optimization algorithm, such as Adam~\cite{kingma2014adam}, which facilitates the iterative updates to both $M_{surr}$
and ${\mathbf{C}}(r)_{dummy}$.
The steps of updating $M_{surr}$
and ${\mathbf{C}}(r)_{dummy}$ are detailed in \autoref{alg:SplitNeRFAttack}.

Once the attacker has successfully trained a surrogate model as detailed in \autoref{alg:SplitNeRFAttack}, it can integrate it with the server model to assemble a complete NeRF model. 
This consolidated model enables the attacker to render the client's scene from any camera pose.

\subsection{\finetuneattack} \label{subsec:attack_advanced}
In this section, we explore a scenario where the attacker gains access to a limited amount of scene image data. 
Research has shown that fine-tuning a surrogate model with a small fraction of leaked data can significantly boost its performance~\cite{fu2022label, xie2023label}. 
With the prevalent use of social networks, acquiring a few scene images has become progressively more accessible for potential attackers. 
However, accurately inferring the camera poses linked to these images poses a challenge. 
This hampers the attacker's ability to effectively utilize the leaked images for developing the surrogate model, as precise pose data is essential for accurate NeRF training.
Next, we clarify the threat model of \finetuneattack and introduce the specific attack methodology.

\subsubsection{Threat Model}
The difference between the threat models of \finetuneattack and \reconstructattack is that \finetuneattack has access to auxiliary information, which is a limited amount of scene image data. 
However, the corresponding camera poses for these images are not available.

\subsubsection{Attack Methodology}
Upon executing the \reconstructattack as elaborated in \autoref{subsec:ModelCompletionAttack}, the attacker can utilize the surrogate model to generate and render 2D scene images captured from different camera poses. 
If the \reconstructattack is effective, the images synthesized by the surrogate model can guide the attacker in deducing the actual poses of the scene images by comparing the rendered images with the real ones. 
Successful alignments between the images and their poses empower the attacker to refine the surrogate model.
Therefore, the key lies in identifying the pose of the leaked images. 
We introduce our pose estimation method to address this challenge.

The attacker can perform a grid search method of camera poses to pinpoint the one that most closely matches the leaked image. 
This process involves dissecting the pose into two components: the 3D position and the camera viewing direction. 
Below, we delineate the search spaces for both components:

\mypara{Search Space of 3D locations}
NeRF typically normalizes three coordinate values within a 3D location to range between $[-1,1]$~\cite{mildenhall2020nerf}. Consequently, the search space for 3D locations spans $[-1,1]^3$. 
If the attacker's search granularity is $x$, it traverse all three coordinates from $[-1, -1+x, \ldots, 1]$. 
This results in the attacker evaluating $(\lceil \frac{2}{x} \rceil +1)^3$ potential 3D locations to find the one closest to the leaked image.

\mypara{Search Space of Viewing Directions}
For any 3D location, an omnidirectional view necessitates traversing 2$\pi$ radians for the horizontal direction and $\pi$ radians for the vertical direction. 
This traversal is typically guided by the Horizontal Field of View (HFOV) and Vertical Field of View (VFOV) of the camera used in the NeRF training. 
Denoting the camera's HFOV as $h$ and VFOV as $v$, the attacker uses polar and azimuth angles in the spherical coordinate system to define viewing directions. 
Hence, the search space for viewing directions spans $[0,2\pi] \times [0, \pi]$. At each 3D location, the attacker incrementally adjusts the azimuth angles by $[0, h, \ldots, 2\pi]$ and polar angles by $[0, v, \ldots, \pi]$ to identify the viewing direction that aligns closest with the leaked image. 
The total number of viewing directions evaluated per location is $(\lceil \frac{2 \pi}{h} \rceil +1)* (\lceil \frac{\pi}{v} \rceil +1)$.

For a leaked image, the attacker typically conducts at most $(\lceil \frac{2}{x} \rceil +1)^3 * (\lceil \frac{2 \pi}{h} \rceil +1)* (\lceil \frac{\pi}{v} \rceil +1)$ iterations to identify a camera pose.
For instance, if NeRF training utilizes an Intel RealSense Depth Camera D435i\footnote{\url{https://www.intelrealsense.com/depth-camera-d435i}} with an HFOV of $69^\circ$ and a VFOV of $42^\circ$, and the granularity of the attack is set to $0.1$, the total number of pose searches reaches approximately $388,962$. 
After initially identifying a pose, the attacker can refine their search using finer granularity for more precise adjustments and potentially better alignment with the target pose. 
For example, employing a granularity of $0.01$ for 3D locations and $1^\circ$ for viewing directions enhances the precision of the pose adjustments.
The attacker uses fidelity metrics, as detailed in \autoref{subsec:attack_setup}, to gauge the closeness of the pose to the leaked image. 
This enables the automated identification of the pose that most precisely corresponds to the leaked image.

\mypara{Discussion}
The challenge of pose estimation is closely related to the inverting NeRF (iNeRF) problem, as discussed by Yen et al.~\cite{yen2021inerf}. 
This problem involves deducing the camera pose relative to a 3D object or scene from a specific image. 
However, the images rendered by the attacker's model in our scenario differ from standard color images, potentially reducing the effectiveness of the iNeRF's optimization process, which is designed for full-color scenarios.
We leave the integration of iNeRF for more efficient pose estimation of leaked images as a direction for future research.

\section{Empirical Evaluation for Privacy Risks} \label{sec:attack_result}

In this section, we first assess the effectiveness of \reconstructattack on the \splitnerf framework, revealing significant privacy vulnerabilities within its architecture. 
Second, we execute the \finetuneattack,  leveraging a limited amount of leaked training image data to improve the fidelity of scene restoration. Lastly, we conduct ablation studies to investigate the impact of various attack strategies on the overall effectiveness of the attacks.

\subsection{Experimental Setup} \label{subsec:attack_setup}

\mypara{Datasets}
We conduct our attack experiments using three well-established public NeRF datasets: the synthetic indoor dataset 3D-FRONT~\cite{fu20213dfront}, Hypersim~\cite{roberts2021hypersim}, and the real indoor dataset ScanNet~\cite{dai2017scannet}.
3D-FRONT provides an extensive collection of large-scale synthetic indoor scenes with intricate room layouts and textured furniture models.
Hypersim is a realistic synthetic dataset tailored for indoor scene understanding, featuring a variety of rendered objects with 3D semantic annotations.
ScanNet is a widely used real-world dataset primarily utilized for indoor 3D object detection, comprising over 1,500 scenes.
Given the potential for indoor scenes to inadvertently disclose sensitive information, these datasets are deliberately chosen to enable a comprehensive assessment of privacy vulnerabilities within the \splitnerf framework.

\mypara{Fidelity Metrics}
In place of visual contrasts between the attack-generated images and the original NeRF-rendered images, we introduce quantitative metrics to evaluate the attack's efficacy.
The attacker can generate both depth and color views from any camera pose using the surrogate model,
making it feasible to evaluate the privacy breach based on the quality of these synthetic views.
The Learned Perceptual Image Patch Similarity (LPIPS)~\cite{zhang2018unreasonable} and Structural Similarity (SSIM)~\cite{wang2004image} metrics are used to assess the perceptual quality of images for privacy leakage evaluation~\cite{huang2021evaluating}.
Therefore, we utilize LPIPS and SSIM to define our fidelity metrics. 
Additionally, we conduct human evaluations to more realistically measure privacy breaches.

\begin{figure*}[!tbp]
\centering
\includegraphics[width=175mm]{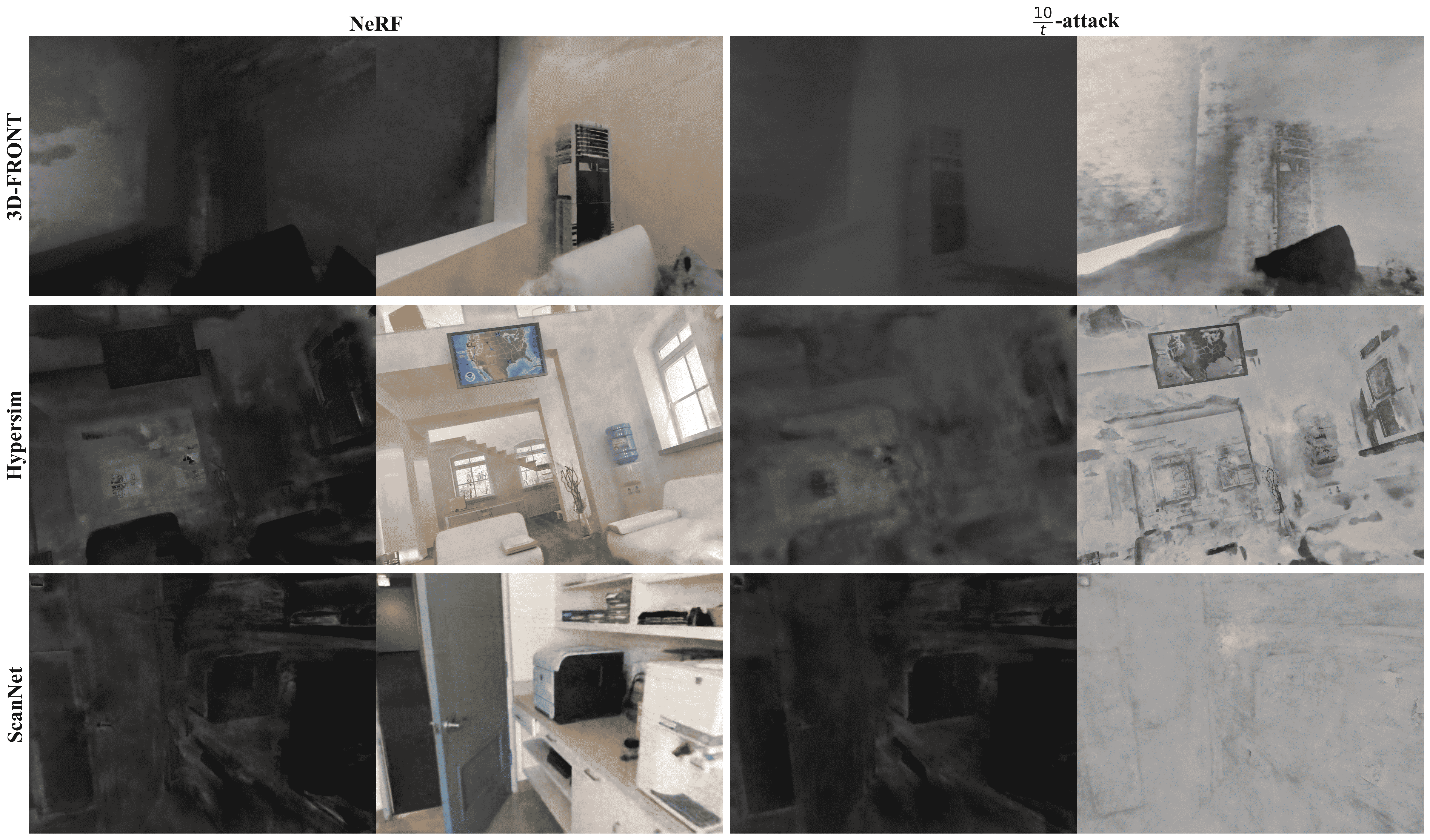}
\caption{\textbf{\reconstructattack results on the three datasets, utilizing $\frac{10}{t}$ learning rate decay schemes, where
$t$ denotes the current learning epoch index.
The attacker successfully restores partial outlines of the actual scenes across all datasets.
Even though the views generated by the attacker are rendered in grayscale, it remains possible to distinguish indoor items and layout scenarios.
The level of detail in the reconstructed scenes highlights a significant privacy violation, as it exposes sensitive information about the structure and contents of the scenes.
}}
\label{fig:attackresult}
\end{figure*}

\begin{itemize}
\item \mypara{LPIPS-depth}
We utilize LPIPS-depth to quantify the perceptual similarity between the depth views rendered by the attacker and those from the original NeRF model.
\item \mypara{LPIPS-gray}
Considering that the attacker's synthetic views are frequently rendered solely in black and white, potentially affecting the synthetical scene's structural assessment, we adjust the original NeRF's color views to grayscale before proceeding with the comparison.
This conversion enables a more targeted evaluation of scene details, decoupled from color influences. Consequently, we use LPIPS-gray to measure the similarity between their grayscale views.

\item \mypara{SSIM-depth}
We utilize SSIM-depth to quantify the structural similarity between the depth views rendered by the attacker and those from the original NeRF model.

\item \mypara{SSIM-gray}
Similarly to LPIPS-gray, we use SSIM-gray to measure the structural similarity between the grayscale views of the attacker and those from the original NeRF model.

\item \mypara{Human-eval}
We present the attacker's views (rendered in depth and color) alongside the real views to the annotators. 
The annotators are then asked to evaluate the severity of privacy disclosure, using a scale from $1$ to $5$, where $1$ indicates no disclosure and $5$ indicates complete disclosure. 
Each view pair is labeled by five independent annotators.
We use Human-eval to report our Human evaluation results.

\end{itemize}

\mypara{NeRF Model Implementation}
We utilize Instant-NGP~\cite{mueller2022instant} through a third-party PyTorch implementation\footnote{A PyTorch CUDA extension implementation of Instant-NGP: \url{https://github.com/ashawkey/torch-ngp}} to implement our experiments. 
This version boosts the efficiency of feature map rendering compared to the original NeRF architecture. 
The model architecture comprises a hash encode network, a two-layer density MLP, and a three-layer color MLP. We employ the Adam optimizer~\cite{kingma2014adam} with an initial learning rate of 0.01, which decays exponentially to $0.1$ of its original size by the final iteration. 
All experiments are executed on a server equipped with two NVIDIA RTX-4090 graphics cards and 256GB of memory.

\mypara{Attack Setting}
\autoref{fig:nerf-attack} depicts the architecture of our attackers. We segment the NeRF network at the two-layer density MLP to specifically isolate output density information locally at the user side, thereby enhancing privacy.
We mirror the client’s model architecture in the attacker’s surrogate model without loss of generality.
This surrogate model consists of a one-layer density MLP and a three-layer color MLP.
To address the issue of potentially disparate scales between the two loss terms,  $L_g$ and $L_{dummy}$,  for the surrogate model, we dynamically adjust 
$\lambda$ to maintain the loss ratio $\frac{\lambda L_g}{L_{dummy}}$ constant.
We assess our attackers using three distinct learning rate decay schemes for training the surrogate model: $0.1^{\frac{t}{T}}, 0.001^\frac{t}{T}, \frac{10}{t}$, where
$T$ is the number of total training epochs and $t$ denotes the current epoch index.

\subsection{\reconstructattack Results}
\label{subsec:attack_result}
In this part, we evaluate the efficacy of \reconstructattack with the loss ratio $\frac{\lambda L_g}{L_{dummy}} = 0.01$ and a learning rate decay scheme of $\frac{10}{t}$,  referred to as the $\frac{10}{t}$-attack.
The experimental results for other attack configurations, involving different loss ratios and decay schemes, are elaborated in \autoref{subsec:attack_ablation} as part of an ablation study aimed at delving deeper into the impact of different configurations on the attack's performance.

\mypara{Experimental Results}
From \autoref{fig:attackresult}, it is evident that the $\frac{10}{t}$-attack method enables partial reconstruction of the original images in terms of both depth and color views, although the color views are presented solely in grayscale.
In conclusion, the $\frac{t}{10}$-attack method can compromise the privacy of the scenes within the three datasets.

In the case of the 3D-FRONT and Hypersim datasets, the attack notably restores recognizable objects in various scenes, such as an air conditioner and sofa in 3D-FRONT, and a water fountain and photo frame in Hypersim. 
This restoration underscores a significant breach of privacy.
In the ScanNet dataset, while the color rendition produced by the attack lacks precision, the depth representation closely mirrors the original scene, enabling object identification. 
These outcomes indicate that the attacker's model has effectively captured the spatial geometry of the scenes.
Despite the synthetic scenes lacking true color fidelity, the attacker's ability to reconstruct scene geometry signifies a substantial compromise in privacy.

\begin{figure*}[!tbp]
\centering
\includegraphics[width=180mm]{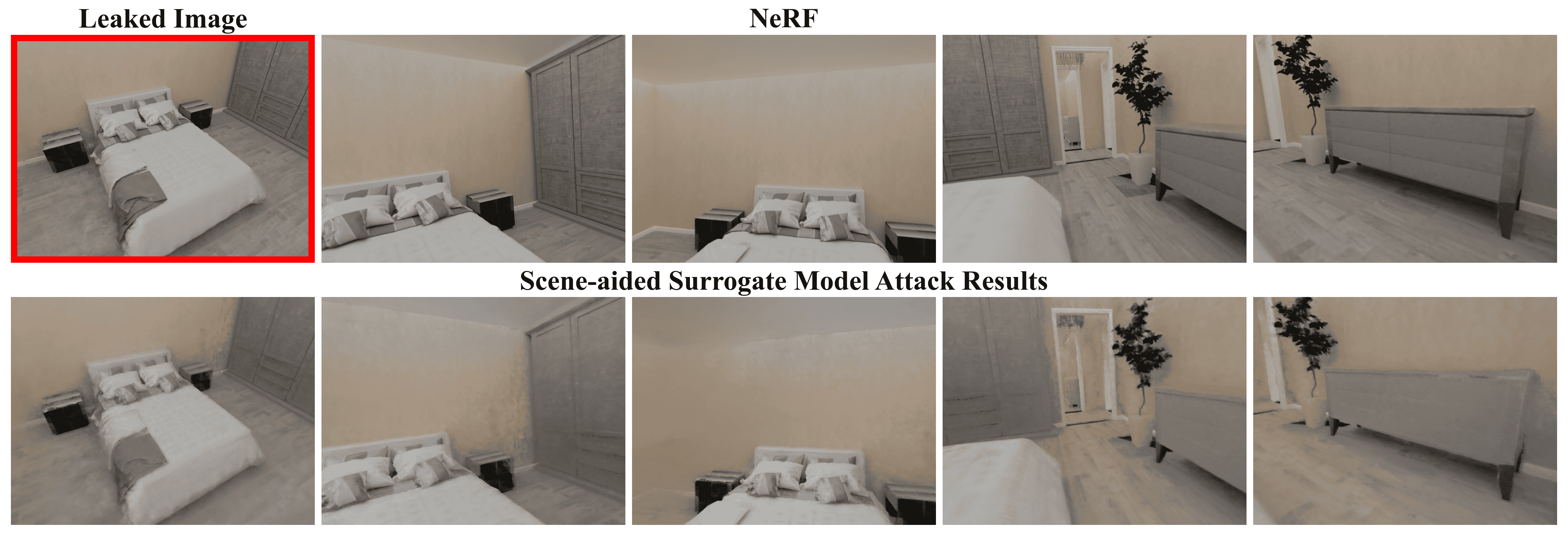}
\caption{\textbf{
\finetuneattack Results.
The results show the attack's ability to significantly restore high-quality, colored views from just one leaked picture, closely resembling the original scene. 
The fidelity of the attack extends beyond the specific pose of the leaked image, accurately replicating various other poses and effectively capturing the entire scene. 
}}
\label{trainprocesseps}
\label{fig:advanced_attack}
\end{figure*}

\subsection{\finetuneattack Results}
\label{subsec:attack_advanced_result}
In this section, we evaluate the effectiveness of the \finetuneattack against the \splitnerf framework, specifically when the attacker has held just a single scene image of the trained scene.

\mypara{Attack Setup}
We execute the \finetuneattack on a bedroom scene from the 3D-FRONT dataset. 
We randomly pick an image as the leaked image from the training data. 
Subsequently, we proceed with the pose estimation method as outlined in \autoref{subsec:attack_advanced}.
This identified pose, alongside the corresponding training data, is then used to fine-tune the attacker's surrogate model in \autoref{subsec:attack_result}.
After \finetuneattack, we render multiple poses from the attack model to visually compare with the original scene in \autoref{fig:advanced_attack}.



\mypara{Results}
As depicted in \autoref{fig:advanced_attack}, the \finetuneattack effectively reconstructs the original scene with a notable Peak Signal to Noise Ratio(PSNR) of $27.62$ and an LPIPS of $0.19$. 
While the details in the scene generated by the attack may appear blurred, and color accuracy may deviate from the actual scene, the overall similarity to the real scene is remarkably close. 
This fidelity is apparent not only in the pose aligned with the leaked image but also in other poses, which closely mirror the original scene.

\mypara{Comparison with \reconstructattack}
The grayscale views in~\autoref{fig:attackresult} resemble the depth views of the scene, which are derived from integrating the spatial density of points.
Since the cut layer output embedding primarily relates to the density output, the attacker acquires the scene's density information more readily.

With the incorporation of the scene image, the attacker gains supervision over color rendering, enabling more effective fine-tuning of the color network. 
Consequently, \finetuneattack more accurately approximates the actual scene, significantly enhancing the fidelity of the reconstructed images.

\subsection{\reconstructattack Ablation Study}\label{subsec:attack_ablation}
We now evaluate the impact of different attack methods concerning the attack performance against \splitnerf.

\mypara{Learning Rate Decay Scheme in \reconstructattack}
The choice of a learning rate decay scheme plays a crucial role in NeRF training, and similarly, it is vital for training the surrogate model in \reconstructattack. 
Initially, we adopt the same learning rate decay strategy used in the original NeRF model training: $0.1^{\frac{t}{T}}$.
Additionally, we introduce two faster decaying schemes: $0.001^\frac{t}{T}, \frac{10}{t}$\footnote{Since $T$ tends to be large in NeRF training, often reaching values like $60,000$, the scheme $\frac{10}{t}$ results in a faster decay rate in practice.}.
We set the loss ratio to be $0.01$.

Quantitative comparison results of these three attack strategies are presented in \autoref{table:attack-different methods}, with visual outcomes provided in Appendix \ref{appendix:attack_ablation}. 
The results suggest that the $\frac{10}{t}$-attack generally outperforms the others. 
During the early stages of training, NeRF primarily focuses on learning geometric information, while later stages are devoted to enhancing color information and details. 
In the initial phases, attackers, who possess part of the geometry network and access gradients situated within this segment, can more effectively acquire geometric information. 
However, in the later stages of training, the gradients returned by the client mainly reflect updates to the color network. 
Since the supervision provided by the surrogate model over the color network updates is not enough, the effectiveness of the attack diminishes during this period.
The $\frac{10}{t}$-attack maintains effective attack performance because it decays more rapidly and seldom updates network parameters in the later stages of training.

\mypara{Surrogate Model's Loss Ratio}
The effectiveness of the surrogate model in our attack framework is governed by the gradient loss $L_g$ and the dummy reconstruction loss $L_{dummy}$.
Consequently, the ratio of these losses, represented as $\frac{\lambda L_g}{L_{dummy}}$, dictates how heavily the surrogate model relies on gradient information during training.

To explore this dependency, we vary the loss ratio from $0.01$ to $100$ for the $\frac{10}{t}$-attack. 
To thoroughly examine the impact of the two loss terms on attacks, we also conduct \reconstructattack focusing solely on $L_g$ (loss ratio $\infty$) or $L_{dummy}$ (loss ratio $0$). 
These attacks are relatively successful, with the human-eval metric generally higher than $4$. 
Detailed visual outcomes are available in \autoref{appendix:attack_ablation}.
Empirical results indicate that all chosen ratios can succeed in scene reconstruction in both depth and color views. 
Considering various attack metrics and visual results, the attack with a loss ratio of $0.01$ performs better.

\mypara{Different Structures of Surrogate Models}
The default structure of the attacker's surrogate model consists of a one-layer density MLP and a three-layer color MLP. 
Variations in surrogate model structure could potentially impact attack performance and defense effectiveness. Therefore, we conduct experiments with two additional surrogate model structures: a one-layer density MLP with a two-layer color MLP, and a one-layer density MLP with a four-layer color MLP. 
These structures represent different surrogate models and attacker capabilities. 
The results show that both structures can successfully attack \splitnerf. 
More detailed results are presented in \autoref{appendix:diff_surrogate_structures}.

\mypara{Discussion about Fidelity Metrics}
The LPIPS score, which measures perceptual similarity, ranges from $0$ to $1$.
A score of $0$ indicates that images are perceptually identical or extremely similar, while a score of $1$ suggests significant perceptual differences.

\begin{table}[!tbp] 
	\centering 
	\caption{\textbf{Comparison of \reconstructattack metrics for different learning rate decay strategies under the loss ratio $0.01$.
    We utilize three distinct strategies: $\frac{10}{t}, 0.1^{\frac{t}{T}},0.001^{\frac{t}{T}}$, where $t$ represents the current iteration and $T$ denotes the total training iterations. 
    We highlight the $\frac{10}{t}$-attack with a \colorbox{pink}{red} background and present the best results in bold.
    $\uparrow(\downarrow)$ means a higher(lower) value is favored.
    }}
        \setlength{\tabcolsep}{0.4mm}
	\begin{tabular}{c c |c c c c c}  
        \toprule
        ~ & ~ & \multicolumn{3}{c}{\textbf{Learning Rate Decay Schemes}} \\
	\textbf{Dataset} & \textbf{Metric} & \textbf{$\frac{10}{t}$-attack} & \textbf{$0.1^{\frac{t}{T}}$-attack} & \textbf{$0.001^{\frac{t}{T}}$-attack} \\ 
        \midrule
        {\multirow{5}*{3D-FRONT}} & LPIPS-depth $\downarrow$ & $\cellcolor{pink} \textbf{0.44}$ & $0.47$ & $0.49$ \\
        ~ & LPIPS-gray $\downarrow$ & $\cellcolor{pink} \textbf{0.49}$ & $0.54$ & $0.54$ \\
        ~ & SSIM-depth $\uparrow$ & $\cellcolor{pink} \textbf{0.77}$ & $0.64$ & $0.62$ \\
        ~ & SSIM-gray $\uparrow$ & $\cellcolor{pink}0.64$ & $\textbf{0.77}$ & $0.77$ \\
        ~ & Human-eval $\uparrow$ & $\cellcolor{pink}{\textbf{4.6}}$ & $1.6$ & $1.6$ \\
         \midrule
        {\multirow{5}*{Hypersim}} & LPIPS-depth $\downarrow$ & $\cellcolor{pink} \textbf{0.39}$ & $0.59$ & $0.58$ \\
        ~ & LPIPS-gray $\downarrow$ & $\cellcolor{pink} \textbf{0.41}$ & $0.65$ & $0.65$ \\
        ~ & SSIM-depth $\uparrow$ & $\cellcolor{pink} \textbf{0.82}$ & $0.22$ & $0.26$ \\
        ~ & SSIM-gray $\uparrow$ & $\cellcolor{pink}0.72$ & $\textbf{0.80}$ & $0.80$ \\
        ~ & Human-eval $\uparrow$ & $\cellcolor{pink}{\textbf{5.0}}$ & $3.2$ & $3.2$ \\
         \midrule
        {\multirow{5}*{ScanNet}} & LPIPS-depth $\downarrow$ & $\cellcolor{pink} \textbf{0.20}$ & $0.22$ & $0.22$ \\
        ~ & LPIPS-gray $\downarrow$ & $\cellcolor{pink} \textbf{0.64}$ & $0.65$ & $0.64$ \\
        ~ & SSIM-depth $\uparrow$ & $\cellcolor{pink} \textbf{0.92}$ & $0.88$ & $0.90$ \\
        ~ & SSIM-gray $\uparrow$ & $\cellcolor{pink}0.58$ & $\textbf{0.60}$ & $0.60$ \\
        ~ & Human-eval $\uparrow$ & $\cellcolor{pink}{\textbf{4.4}}$ & $3.4$ & $3.6$ \\
        \bottomrule
	\end{tabular}   
 \label{table:attack-different methods}
\end{table}

The fidelity metric values exhibit significant variability across various scenarios. 
For instance, despite the visual recovery of many color view contours in the 3D-FRONT dataset in \autoref{fig:attackresult}, its LPIPS-gray value is high at $0.49$ in \autoref{table:attack-different methods}. 
Conversely, the ScanNet dataset demonstrates a successful depth view attack, evidenced by a low LPIPS-depth value of $0.20$. 
These examples highlight that, while the attack outcomes may visually seem effective, the fidelity metrics can vary significantly.
Hence, to accurately assess attack effectiveness, comparisons must be made within the same scene.
This can be further illustrated by examining the visuals and metrics detailed in \autoref{table:attack-different methods}, \autoref{table:attack-loss ratio}, and \autoref{appendix:attack_ablation}.

It is also important to note that attack effectiveness does not uniformly reflect across all fidelity metrics. 
For example, in the ScanNet dataset in \autoref{fig:attackresult}, while the depth attack is successful with a low LPIPS-depth value of $0.20$, the color views do not match this effectiveness, as indicated by higher values of LPIPS-gray. 
A score of around $0.7$ in LPIPS-based metrics indicates an unsuccessful attack, emphasizing that success in one aspect does not guarantee overall effectiveness.

The same conclusion applies to SSIM-based metrics. The SSIM score, which measures structural similarity, ranges from $0$ to $1$. 
A score of $1$ indicates that images are structurally identical or extremely similar, while a score of $0$ suggests significant structural differences. 
In the ScanNet dataset (see \autoref{fig:attackresult}), while the depth attack is successful with a high SSIM-depth value of $0.92$, the color views do not match this effectiveness, as indicated by low SSIM-gray values. 
Overall, the Human-eval metric is closer to the real privacy disclosure situation, although it is subject to human subjectivity.
Specifically, for the results in \autoref{fig:attackresult}, the Human-eval metric remains high, consistently above $4.4$ as shown in \autoref{table:attack-different methods}.

\begin{table}[!tbp] 
	\centering 
	\caption{\textbf{
    Comparison of \reconstructattack metrics for different surrogate model's loss ratio under $\frac{10}{t}$-attack.
    A loss ratio of $0$ indicates attacks solely on $L_{dummy}$, whereas a loss ratio of $\infty$ indicates attacks solely on $L_g$.
    $\uparrow(\downarrow)$ means a higher(lower) value is favored.
 }}
        \setlength{\tabcolsep}{0.8 mm}
	\begin{tabular}{c c | c c c c c c c}  
        \toprule
        ~ & ~ & \multicolumn{7}{c}{\textbf{Loss Ratio}} \\
        \textbf{Dataset} & \textbf{Metric} & $0$ & $0.01$ & $0.1$ & $1$ & $10$ &  $100$ & $\infty$\\ 
        \midrule
        {\multirow{5}*{3D-FRONT}} & LPIPS-depth $\downarrow$ & $0.22$ & $0.44$ & $0.44$ & $0.41$ & $0.44$ & $0.46$ & $0.46$ \\   
        ~ & LPIPS-gray $\downarrow$ & $0.46$ & $0.49$ & $0.50$ & $0.49$ & $0.51$ & $0.50$ & $0.47$ \\   
        ~ & SSIM-depth $\uparrow$ & $0.89$ & $0.77$ & $0.77$ & $0.79$ & $0.78$ & $0.78$ & $0.78$ \\   
        ~ & SSIM-gray $\uparrow$ & $0.70$ & $0.64$ & $0.64$ & $0.60$ & $0.58$ & $0.59$ & $0.63$ \\   
        ~ & Human-eval $\uparrow$ & $4.4$ & $4.6$ & $4.6$ & $4.6$ & $4.4$ & $4.4$ & $4.0$ \\   
        \midrule
        {\multirow{5}*{Hypersim}} & LPIPS-depth $\downarrow$ & $0.13$ & $0.39$ & $0.49$ & $0.52$ & $0.56$ & $0.57$ & $0.56$ \\ 
        ~ & LPIPS-gray $\downarrow$ & $0.47$ & $0.41$ & $0.43$ & $0.46$ & $0.47$ & $0.47$ & $0.48$ \\  
        ~ & SSIM-depth $\uparrow$ & $0.96$ & $0.82$ & $0.78$ & $0.78$ & $0.78$ & $0.78$ & $0.78$ \\   
        ~ & SSIM-gray $\uparrow$ & $0.78$ & $0.72$ & $0.69$ & $0.66$ & $0.65$ & $0.66$ & $0.61$ \\   
        ~ & Human-eval $\uparrow$ & $4.6$ & $5.0$ & $4.4$ & $4.8$ & $4.8$ & $4.8$ & $5.0$\\  
        \midrule
        {\multirow{5}*{ScanNet}} & LPIPS-depth $\downarrow$ & $0.22$ & $0.20$ & $0.44$ & $0.50$ & $0.56$ & $0.63$ & $0.65$ \\
        ~ & LPIPS-gray $\downarrow$ & $0.66$ & $0.64$ & $0.55$ & $0.54$ & $0.55$ & $0.55$ & $0.56$ \\
        ~ & SSIM-depth $\uparrow$ & $0.91$ & $0.92$ & $0.76$ & $0.73$ & $0.73$ & $0.73$ & $0.73$ \\   
        ~ & SSIM-gray $\uparrow$ & $0.54$ & $0.58$ & $0.52$ & $0.51$ & $0.49$ & $0.49$ & $0.53$ \\   
        ~ & Human-eval $\uparrow$ & $4.0$ & $4.4$ & $4.0$ & $4.2$ & $4.2$ & $4.2$ & $4.4$\\  
        \bottomrule
	\end{tabular}
 \label{table:attack-loss ratio}
\end{table}

\section{\securesplitnerf} \label{sec:defense}
Having identified significant privacy breaches within the \splitnerf framework, our next step is to develop a version that ensures privacy. 
The privacy-preserving framework should satisfy the following objectives:

\begin{itemize}
    \item \mypara{(G1) Defense Effectiveness}
    The framework must effectively address the identified privacy vulnerabilities. Specifically, the framework needs to protect against attackers who utilize gradient information and potentially some leaked images.
    
    \item \mypara{(G2) Model Utility}
    It is critical for the framework to maintain the utility of the NeRF model, ensuring it can truthfully and effectively represent the target scenes in 3D reconstruction.
\end{itemize}

\begin{figure*}[!tbp]
\centering
\includegraphics[width=180mm]{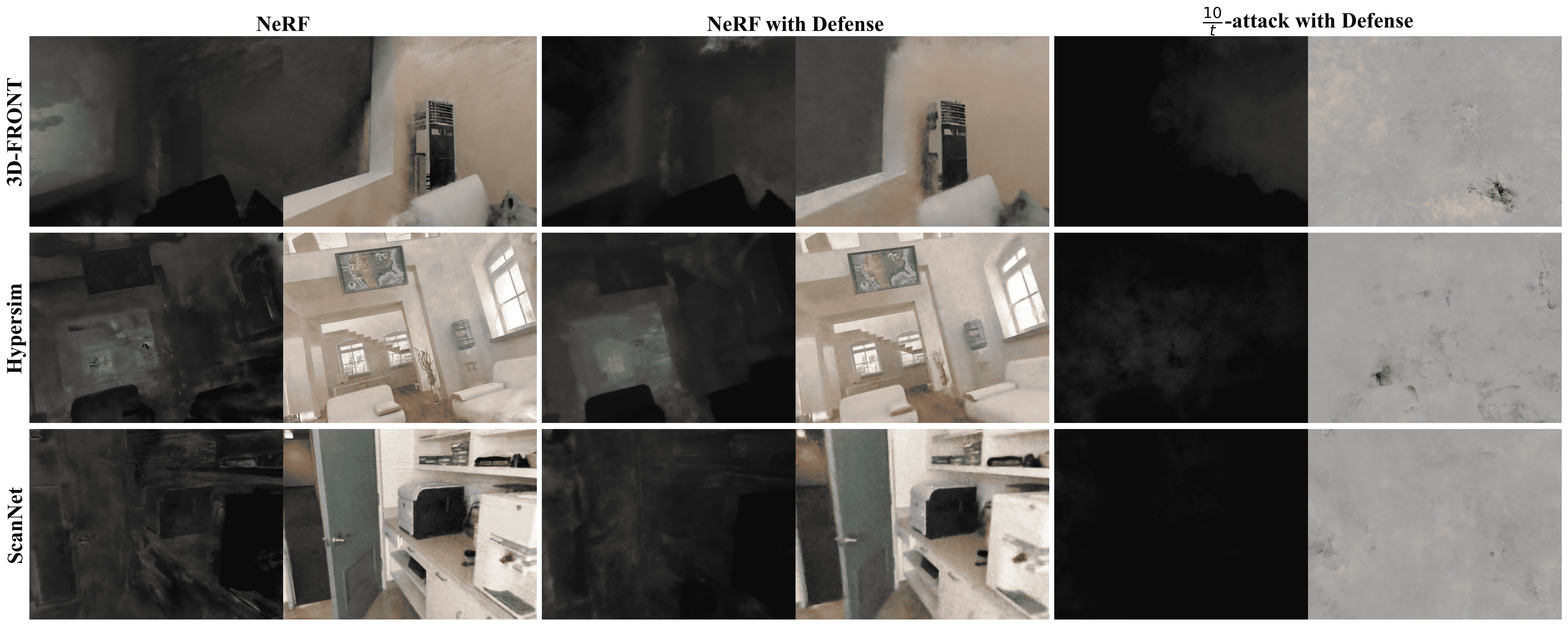}
\caption{\textbf{
The defense results of \securesplitnerf with the configure $c = 1.2, r = 0.0001$ under $\frac{10}{t}$-attack, where $c$ and $r$ denote the noise scale and noise decay ratio.
\securesplitnerf achieves a balanced trade-off between privacy protection and model utility.
In particular, the attacker can not recover any useful information, while the NeRF with defense can still reflect the appearance of the scene, albeit with some noise in the views.
}}
\label{fig:defense-10-k}
\end{figure*}

\mypara{\labelnoisenerf}
Some prevalent defense methods against malicious attacks in SL involve protecting private label information by adding noises to the labels~\cite{ghazi2021deep, qiu2023defending}.  
In a NeRF network, labels are the density and color values of points in 3D space, which are implicit and must be integrated to determine the pixel colors.
We introduce a basic solution \labelnoisenerf, which protects privacy by adding noises to the image pixels.
\labelnoisenerf can be outlined as follows: for each ray $r$, whose RGB ground truth value is $\mathbf{C}(r)$, we apply
$$
\begin{gathered}
\left\langle \mathbf{C}(r) \right\rangle = \mathbf{C}(r) + \xi,\ \xi \sim \mathcal{N}(\boldsymbol{0}, \sigma_{l}^2),
\end{gathered}
$$  
where $\mathcal{N}(\boldsymbol{0}, \sigma_{l}^2)$ represents a Gaussian random variable with variance $\sigma_{l}$. 
This altered RGB value 
$\left\langle \mathbf{C}(r) \right\rangle$ is then utilized in the loss function during NeRF training, enabling successful backpropagation.
While \labelnoisenerf safeguards the privacy of the image data, the model's privacy during NeRF training may remain vulnerable. The introduction of noise in the image dataset can obscure the network parameters, but it also unavoidably diminishes the model's utility.
Moreover, it may not be able to defend against attacks that exploit gradient sharing. 
Consideration of more advanced privacy-preserving methods is essential to address these limitations and vulnerabilities.

\mypara{\securesplitnerf}
The \securesplitnerf framework entails obfuscating gradient information, which malicious entities could potentially leverage~\cite{li2021label, zhu2019deep}. 
This is achieved by introducing randomly generated noise, including Gaussian or Laplace distributed ones, to obscure the genuine gradient, thereby impeding access to the valuable information during the gradient sharing~\cite{mohammady2020r2dp, xie2023label, abadi2016deep, huang2024differential, yang2022differentially}.
In the defense mechanism, the noise applied to the gradients diminishes in intensity relative to the training iteration index, with the noise scale decreasing exponentially as training advances. 
In our algorithm, Gaussian noise is introduced to the shared gradients of each training batch $g_1, \ldots, g_B$, where $B$ represents the batch size, in the following manner:
$$
\begin{gathered}
\left\langle g_i \right\rangle = g_i + \eta_i, \ \eta_i \sim \mathcal{N}(\boldsymbol{0}, \sigma_t^2 I_d),
\end{gathered}
$$
Here, $\sigma_t = c \max_{i=1,\ldots,B} \lVert g_i \rVert r ^{\frac{t}{T}}$, where $d$ denotes the dimension of the gradients and $I_d$ is the $d$-by-$d$ identity matrix. The magnitude of $\sigma_t$ is regulated as follows: the coefficient $c$ acts as a scaling factor determining the noise magnitude, while $r$ governs the decay rate.
Consequently, the noise intensity not only diminishes as training progresses but also adjusts to the volume of the gradients. This adaptive approach enables the noise addition to dynamically adapt to various training scenarios.

In the realm of deep learning, it has been established that under certain conditions, decaying noise does not harm the eventual convergence of neural network training~\cite{ward2020adagrad}. 
Moreover, our attack experiments have unveiled that the surrogate model has the capacity to absorb significant information in the initial phases of training, emphasizing the crucial necessity for intensive noise injection during these early stages.

\mypara{Different Noises' Influence}
In \securesplitnerf, the noise only needs to satisfy zero mean, with variance proportional to the norm of the gradient. 
In the domain of privacy-preserving deep learning, Gaussian noise is commonly used to meet this requirement~\cite{abadi2016deep, wang2024dpadapter}.

\subsection{Algorithm Analysis}
In this section, we first theoretically analyze the communication and computation complexity of \securesplitnerf, and then empirically evaluate the communication overhead and running time. Recall that the dimension of the cut layer is $d$. We denote the number of parameters of the client model as $P$, and the total training iterations as $T$.

\mypara{Communication Complexity}
The communication cost can be broken as:
(1) the user uploads the sampled $N_p$ points along $N_r$ rays to the server($O(T N_r N_p)$ messages);
(2) the server shares the cut layer embeddings with the user($O(T N_r N_p d)$ messages);
(3) The user returns the cut layer gradients to the server($O(T N_r N_p d)$ messages).
Hence, the total communication overhead is $O(T N_r N_p d)$.

\mypara{User Computation Complexity}
The user’s computation cost can be broken as: 
(1) forward propagation and backpropagation computation of the client model($O(T N_r N_p P)$ complexity);
(2) perform the noise adding($O(T N_r N_p d)$ complexity).
Therefore, the user’s computation cost adds up to $O(T N_r N_p P)$.

\mypara{Empirical Evaluation}
For our main evaluations, we adopt the default setting of Instant-NGP~\cite{mueller2022instant} as the standard version. 
Specifically, the standard version uses $N_r = 4096$ and $N_p = 512$, with the client model consisting of a one-layer density MLP followed by a three-layer color MLP, each with hidden dimensions of 64.
We use the standard version to better validate \splitnerf's privacy issues and \securesplitnerf's security and utility. 
To reduce communication and computation stress, we design a light version of \securesplitnerf.
The light version uses $N_r = 512$ and $N_p = 128$, with the client model consisting of a one-layer density MLP followed by a two-layer color MLP, each with hidden dimensions of 32. 
The communication overhead of a single iteration is 8 Megabytes. 
The user running time per iteration is 0.021, 0.019, and 0.021 seconds on the 3D-FRONT, Hypersim, and ScanNet datasets, respectively.
We show in \autoref{appendix:light_results} that the light version \securesplitnerf still provides stable defense effectiveness and acceptable utility.


\section{Empirical Evaluation for \securesplitnerf } \label{sec:defense_result}

In this section, we first assess the defense effectiveness and model utility of \securesplitnerf. 
Second, we conduct ablation studies to explore how various configurations of \securesplitnerf influence both defense effectiveness and model utility.

\begin{table*}[!tbp] 
	\centering 
	\caption{\textbf{\securesplitnerf ablation study under $\frac{10}{t}$-attack with $0.01$ loss ratio.
        We evaluate different noise scales $c$ and decay ratio $r$.
        Experimental results indicate that \securesplitnerf with $c= 1.2, r = 0.0001$
        achieves a better trade-off between privacy and utility.
        We use PSNR and LPIPS to compare utility and use LPIPS-depth, LPIPS-gray, SSIM-depth, SSIM-gray, and Human-eval to measure privacy.
        We report the utility results without defense, using GT as the baseline.
        We highlight $c= 1.2, r = 0.0001$ method in the \colorbox{pink}{red} ground.
        $\uparrow(\downarrow)$ means a higher(lower) value is favored.}}
        \setlength{\tabcolsep}{0.5 mm}
	\begin{tabular}{c c | c c c c c c c c c c c c}  
        \toprule
        ~ & ~ & \multicolumn{12}{c}{\textbf{\securesplitnerf Configuration}} \\
        {\multirow{2}*{\textbf{Dataset}}} & {\multirow{2}*{\textbf{Metric}}} & $c = 0.6$ & $c =0.6$ & $c =0.6$ & $c = 1.2$ & $c = 1.2$ & $c = 1.2$ & $c = 2.4$ & $c = 2.4$ & $c = 2.4$ & $c = 4.8$ & $c = 4.8$ & {\multirow{2}*{\textbf{GT}}} \\ 
        ~ & ~ & $r = 0.0001$ & $r = 0.001$ & $r = 1$ & $r = 0.0001$ & $r = 0.001$ & $r = 1$ & $r = 0.0001$ & $r = 0.001$ & $r = 1$ &  $r = 0.0001$ & $r = 0.001$ &  \\
        \midrule
        {\multirow{7}*{3D-FRONT}} & LPIPS-depth $\uparrow$ & $0.30$ & $0.35$ & $0.50$ & $\cellcolor{pink}0.42$ & $0.43$ & $0.60$ & $0.42$ & $0.45$ & $0.56$ & $0.52$ & $0.52$ & - \\
        ~ & LPIPS-gray $\uparrow$ & $0.46$ & $0.49$ & $0.44$ & $\cellcolor{pink}0.47$ & $0.50$ & $0.45$ & $0.58$ & $0.50$ & $0.46$ & $0.45$ & $0.52$ & - \\ 
        ~ & SSIM-depth $\downarrow$ & $0.83$ & $0.81$ & $0.59$ & $\cellcolor{pink}0.53$ & $0.75$ & $0.47$ & $0.80$ & $0.71$ & $0.40$ & $0.13$ & $0.55$ & - \\ 
        ~ & SSIM-gray $\downarrow$ & $0.74$ & $0.73$ & $0.75$ & $\cellcolor{pink}0.72$ & $0.73$ & $0.74$ & $0.63$ & $0.72$ & $0.74$ & $0.74$ & $0.70$ & - \\ 
        ~ & Human-eval $\downarrow$ & $2.0$ & $1.6$ & $1.0$ & $\cellcolor{pink}1.0$ & $1.0$ & $1.0$ & $1.0$ & $1.0$ & $1.0$ & $1.0$ & $1.0$ & - \\ 
        ~ & PSNR $\uparrow$ & $21.04$ & $20.95$ & $19.70$ & $\cellcolor{pink}20.83$ & $20.82$ & $18.13$ & $20.64$ & $20.25$ & $17.32$ & $20.00$ & $19.42$ & $21.24$ \\
        ~ & LPIPS $\downarrow$ & $0.42$ & $0.47$ & $0.71$ & $\cellcolor{pink}0.45$ & $0.49$ & $0.77$ & $0.51$ & $0.61$ & $0.80$ & $0.65$ & $0.67$ & $0.39$ \\
        \midrule
        {\multirow{7}*{Hypersim}} & LPIPS-depth $\uparrow$ & $0.41$ & $0.51$ & $0.52$ & $\cellcolor{pink}0.51$ & $0.54$ & $0.60$ & $0.60$ & $0.60$ & $0.66$ & $0.61$ & $0.67$ & -\\
        ~ & LPIPS-gray $\uparrow$ & $0.55$ & $0.56$ & $0.56$ & $\cellcolor{pink}0.56$ & $0.57$ & $0.54$ & $0.67$ & $0.71$ & $0.73$ & $0.70$ & $0.74$ & -\\
        ~ & SSIM-depth $\downarrow$ & $0.84$ & $0.63$ & $0.77$ & $\cellcolor{pink}0.63$ & $0.55$ & $0.69$ & $0.79$ & $0.79$ & $0.76$ & $0.79$ & $0.75$ & -\\ 
        ~ & SSIM-gray $\downarrow$ & $0.79$ & $0.79$ & $0.79$ & $\cellcolor{pink}0.78$ & $0.79$ & $0.79$ & $0.71$ & $0.70$ & $0.65$ & $0.70$ & $0.64$ & -\\ 
        ~ & Human-eval $\downarrow$ & $2.0$ & $1.0$ & $1.0$ & $\cellcolor{pink}1.0$ & $1.0$ & $1.0$ & $1.0$ & $1.0$ & $1.0$ & $1.0$ & $1.0$ & - \\  
        ~ & PSNR $\uparrow$ &$18.69$ & $18.65$ & $18.14$ & $\cellcolor{pink}18.66$ & $18.59$ & $17.79$ & $18.64$ & $18.53$ & $17.79$ & $18.59$ & $18.20$ & $18.77$ \\
        ~ & LPIPS $\downarrow$ & $0.44$ & $0.49$ & $0.75$ & $\cellcolor{pink}0.48$ & $0.59$ & $0.81$ & $0.54$ & $0.67$ & $0.81$ & $0.61$ & $0.76$ & $0.36$ \\
        \midrule
        {\multirow{7}*{ScanNet}} & LPIPS-depth $\uparrow$ & $0.54$ & $0.55$ & $0.62$ & $\cellcolor{pink}0.57$ & $0.56$ & $0.56$ & $0.59$ & $0.59$ & $0.58$ & $0.55$ & $0.58$ & -\\
        ~ & LPIPS-gray $\uparrow$ & $0.69$ & $0.68$ & $0.65$ & $\cellcolor{pink}0.67$ & $0.67$ & $0.64$ & $0.71$ & $0.65$ & $0.63$ & $0.67$ & $0.66$ & -\\
        ~ & SSIM-depth $\downarrow$ & $0.71$ & $0.54$ & $0.30$ & $\cellcolor{pink}0.49$ & $0.59$ & $0.44$ & $0.43$ & $0.38$ & $0.34$ & $0.43$ & $0.42$ & -\\ 
        ~ & SSIM-gray $\downarrow$ & $0.60$ & $0.60$ & $0.60$ & $\cellcolor{pink}0.60$ & $0.60$ & $0.60$ & $0.61$ & $0.59$ & $0.59$ & $0.60$ & $0.60$ & -\\ 
        ~ & Human-eval $\downarrow$ & $1.2$ & $1.0$ & $1.0$ & $\cellcolor{pink}1.0$ & $1.0$ & $1.0$ & $1.0$ & $1.0$ & $1.0$ & $1.0$ & $1.0$ & - \\
        ~ & PSNR $\uparrow$ & $19.63$ & $19.51$ & $17.94$ & $\cellcolor{pink}19.59$ & $19.44$ & $17.48$ & $19.51$ & $19.22$ & $17.08$ & $19.42$ & $18.81$ & $19.89$ \\
        ~ & LPIPS $\downarrow$ & $0.52$ & $0.58$ & $0.80$ & $\cellcolor{pink}0.56$ & $0.62$ & $0.82$ & $0.60$ & $0.72$ & $0.82$ & $0.67$ & $0.77$ & $0.42$ \\
        \bottomrule
	\end{tabular}
 \label{table:defense_ablation}
\end{table*}

\subsection{Experimental Setup} \label{subsec:defense_setup}

The experimental setup for implementing the NeRF model and the datasets remains consistent with the details provided in \autoref{subsec:attack_setup}.

\mypara{Defense Metrics}
In the design of \securesplitnerf, we mainly consider two aspects of performance, defense effectiveness and model utility.

\begin{itemize}
\item \mypara{Defense Effectiveness}
We use the fidelity metrics(LPIPS-depth, LPIPS-gray, SSIM-depth, SSIM-color) in \autoref{subsec:attack_setup} to measure the decline of the attack effect, and thus to reflect the defense effectiveness.

\item \mypara{Model Utility}
We use PSNR and LPIPS to measure the NeRF model utility of \securesplitnerf.
\end{itemize}

\mypara{Defense Setting}
To evaluate the defense robustness of \securesplitnerf, we expose it to three different attacks using varying learning rate schemes ($\frac{10}{t}$, $0.1^{\frac{t}{T}}$, $0.001^\frac{t}{T}$) and a loss ratio of $0.01$.
We specifically showcase the $\frac{10}{t}$-attack in \autoref{subsec:defense_results} and \autoref{subsec:defense_ablation}, because our experiments in \autoref{subsec:attack_ablation} confirm it as the most potent attack configuration. 
The results of the $0.1^{\frac{t}{T}}$-attack and $0.001^{\frac{t}{T}}$-attack are similar to the $\frac{10}{t}$-attack's result, and these are detailed in \autoref{appendix:defense_ablation}.

\subsection{Effectiveness} \label{subsec:defense_results}

\mypara{Results}
The defense outcomes of \securesplitnerf in \autoref{fig:defense-10-k} are configured with $c = 1.2, r = 0.0001$. 
We observed a significant reduction in the attack's effectiveness: the images generated by the attacker contained little to no useful information.
The defense effectiveness is more obvious when compared to the successful attack depicted in \autoref{fig:attackresult}.

Regarding model utility, the quality of NeRF generation with \securesplitnerf does experience a slight reduction. 
However, this decrease remains within acceptable limits. 
While the clarity of object boundaries may be somewhat diminished, the overall structure of the scene is still rendered clearly, preserving the essential visual integrity of the model.

\mypara{Optimal Defense Setting}
An ideal defense setting should demonstrate the capability to thwart attacks while maintaining acceptable model utility effectively. Additionally, the defense parameters should be versatile and applicable across various scenarios.

In \autoref{fig:defense-10-k}, we confirm that this defense setting provides strong defensive effectiveness and maintains good model utility across three distinct datasets. 
Despite variations in the parameters of the NeRF network models trained in different scenarios, 
\securesplitnerf's reliance on gradient norms enhances its generalizability.

\mypara{Defense for \finetuneattack}
In \autoref{subsec:attack_advanced}, the success of \finetuneattack is heavily dependent on the effectiveness of the preceding \reconstructattack. 
\finetuneattack specifically needs a successful \reconstructattack model to accurately determine the poses closest to the leaked images.
Our defense results demonstrate a robust capability to neutralize the \reconstructattack, making it impossible to extract any useful information. 
With depth views rendered entirely black and color views appearing gray, \reconstructattack is unable to provide the necessary insights for \finetuneattack to function effectively.
Consequently, \finetuneattack is ineffective under this condition, affirming that \securesplitnerf provides a strong defense against both the \reconstructattack and \finetuneattack.

\subsection{\securesplitnerf Ablation Study} \label{subsec:defense_ablation}
In this section, we evaluate the performances of various defense configurations within the \securesplitnerf framework. 
Our objective is to identify the defense settings that offer the best balance between robust defense effectiveness and acceptable model utility.

\mypara{\securesplitnerf Results}
For \securesplitnerf, we explore noise scales $c$ at set $\{0.6, 1.2,2.4,4.8\}$ and decay ratio $r$ at $\{0.0001, 0.001, 1\}$, where $r = 1$ indicates adding noise with no decay. 
The quantitative results are documented in \autoref{table:defense_ablation}, with visual results available in \autoref{appendix:defense_ablation}.

Through our evaluations, we find that $c = 1.2, r = 0.0001$ offers a balanced defense across all three datasets. 
This setting provides better model utility compared to the non-decaying defense mechanism($r = 1$).
A lower noise scale though yielding higher model utility, results in considerable privacy compromises, which are unacceptable. 
For example, while the $c = 0.6, r = 0.0001$ setting achieves a PSNR of $21.04$ in 3D-FRONT dataset, it also allows the attacker to reach an SSIM-depth of $0.83$, indicating a significant privacy breach.

Conversely, the setup of $c = 1.2, r = 0.0001$ maintains both high utility and robust defense effectiveness. 
Specifically, this setup achieves a PSNR of 20.83 in 3D-FRONT, a decrease of only $2\%$ from the ground truth of 21.24, and provides a strong privacy guarantee, as indicated by a low SSIM-depth value of 0.53.
On the other hand, a higher noise scale($c = 2.4$) enhances defense effectiveness but at the cost of reduced model utility. 
An intuitive comparison of these effects is detailed in \autoref{appendix:defense_ablation}.


\mypara{\labelnoisenerf Results}
For \labelnoisenerf, we explore varying the noise scale $\sigma_l$ at levels of $\{0.5,1,2,4,8\}$.
The quantitative results are presented in \autoref{table:defense_ablation}, and the visual results are available in \autoref{appendix:defense_ablation}.

Overall, \labelnoisenerf has proven to be ineffective. 
Even with high noise levels, it fails to provide a stable defense against attacks.
For instance, at a noise scale of $\sigma_l = 8$, although the model utility significantly decreases (PSNR of $19.17$ and LPIPS of $0.68$ on 3D-FRONT), the attacker's LPIPS-depth remains low at $0.29$, indicating that the defense does not sufficiently obscure the information from the attacker. 
More detailed visual effects demonstrating these outcomes can be viewed in \autoref{appendix:strawman_sol}. 

\begin{table}[!tbp] 
	\centering 
	\caption{\textbf{\labelnoisenerf results under $\frac{10}{t}$-attack with $0.01$ loss ratio.
    We report the utility results without defense, using GT as the baseline.    
     \labelnoisenerf proves ineffective: it fails to defend against attacks when the noise level is high and significantly compromises model utility.
     $\uparrow(\downarrow)$ means a higher(lower) value is favored.
    }}
        \setlength{\tabcolsep}{0.8 mm}
	\begin{tabular}{c c|c c c c c c}  
        \toprule
        ~ & ~ & \multicolumn{5}{c}{\textbf{The Scale of the Noise}} & {\multirow{2}*{\textbf{GT}}}\\
        \textbf{Dataset} & \textbf{Metric}&$0.5$&$1$&$2$&$4$& $8$ \\ 
        \midrule
        {\multirow{7}*{3D-FRONT}} & LPIPS-depth $\uparrow$ & $0.25$ & $0.19$ & $0.20$ & $0.25$ & $0.29$ &-\\ 
        ~ & LPIPS-gray $\uparrow$ & $0.41$ & $0.47$ & $0.42$ & $0.48$ & $0.48$ &-\\ 
        ~ & SSIM-depth $\downarrow$ & $0.89$ & $0.95$ & $0.94$ & $0.90$ & $0.90$ & -\\
        ~ & SSIM-gray $\downarrow$ & $0.76$ & $0.73$ & $0.76$ & $0.74$ & $0.73$ &-\\ 
        ~ & Human-eval $\downarrow$ & $4.0$ & $3.8$ & $3.6$ & $3.0$ & $2.0$ &-\\ 
        ~ & PSNR $\uparrow$ & $20.48$ & $19.93$ & $20.42$ & $20.14$ & $19.17$ &  $21.24$\\
        ~ & LPIPS $\downarrow$ & $0.40$ & $0.47$ & $0.51$ & $0.64$ & $0.68$ & $0.39$\\
        \midrule
        {\multirow{5}*{Hypersim}} & LPIPS-depth $\uparrow$ & $0.22$ & $0.25$ & $0.33$ & $0.42$ & $0.52$ &-\\
        ~ & LPIPS-gray $\uparrow$ & $0.60$ & $0.58$ & $0.55$ & $0.57$ & $0.62$ &-\\ 
        ~ & SSIM-depth $\downarrow$ & $0.93$ & $0.92$ & $0.89$ & $0.86$ & $0.80$ &-\\
        ~ & SSIM-gray $\downarrow$ & $0.80$ & $0.80$ & $0.80$ & $0.79$ & $0.76$ &-\\ 
        ~ & Human-eval $\downarrow$ & $4.0$ & $3.8$ & $3.6$ & $3.0$ & $2.2$ &-\\ 
        ~ & PSNR $\uparrow$ & $18.69$ & $18.72$ & $18.66$ & $18.66$ & $18.48$ & $18.77$\\
        ~ & LPIPS $\downarrow$ & $0.41$ & $0.47$ & $0.56$ & $0.69$ & $0.80$ & $0.36$\\
        \midrule
        {\multirow{5}*{ScanNet}} & LPIPS-depth $\uparrow$ & $0.29$ & $0.33$ & $0.43$ & $0.47$ & $0.44$ &-\\
        ~ & LPIPS-gray $\uparrow$ & $0.65$ & $0.65$ & $0.69$ & $0.70$ & $0.74$ &-\\ 
        ~ & SSIM-depth $\downarrow$ & $0.91$ & $0.90$ & $0.84$ & $0.80$ & $0.81$ &-\\
        ~ & SSIM-gray $\downarrow$ & $0.59$ & $0.59$ & $0.60$ & $0.58$ & $0.53$ &-\\ 
        ~ & Human-eval $\downarrow$ & $5.0$ & $5.0$ & $3.8$ & $3.0$ & $1.0$ &-\\ 
        ~ & PSNR $\uparrow$ & $19.71$ & $19.58$ & $19.48$ & $19.50$ & $18.76$ & $19.89$ \\
        ~ & LPIPS $\downarrow$ & $0.50$ & $0.55$ & $0.61$ & $0.65$ & $0.75$ & $0.42$\\
        \bottomrule
	\end{tabular}
 \label{table:defense_ablation_label_noise}
\end{table}

\section{Discussion} \label{sec:discuss}
\mypara{Privacy Assessment on Reconstructed Images}
The \reconstructattack aligns with the goals of traditional reconstruction attacks that intercept gradients from a model to recreate images, potentially exposing sensitive details of the original images~\cite{fredrikson2015model, zhu2019deep}. 
Our evaluation assesses whether these reconstructed images compromise the privacy of the original scenes, a concern highlighted in recent studies~\cite{sun2024privacy}. 
Common metrics like PSNR, LPIPS, and structural similarity index(SSIM) are traditionally used to measure pixel-level image similarity. 
High similarity scores typically indicate a successful reconstruction attack and greater model vulnerability to privacy breaches. 
Conversely, low scores suggest reduced privacy risks. 
However, the effectiveness of these metrics in accurately reflecting human perceptions of privacy is questionable due to the subjective nature of privacy~\cite{sun2024privacy}.

To address these challenges in NeRF scenarios, we introduce two fidelity metrics based on LPIPS: LPIPS-depth, and LPIPS-gray; and two fidelity metrics based on SSIM: SSIM-depth, and SSIM-gray.
These metrics aim to better align with human perception, though they may still contain inaccuracies.
At the same time, we also conduct human evaluations, adding the Human-eval metric to assess the level of privacy disclosure more realistically.

\mypara{The Trade-off Between Utility and Privacy}
Balancing utility (model performance) and privacy is crucial in privacy-enhancing technologies. 
A key challenge is how to effectively add noise for privacy without sacrificing training performance. 
Our defense approach, which incorporates a noise decay mechanism, offers a better trade-off than traditional noise injection, yet there is potential for further improvement. 
Previous strategies have aimed to mitigate the impact of noise on model utility~\cite{papernot2016semi, tramer2020differentially, yu2021not, zhu2020private, wang2024dpadapter}. 
However, such methods are less effective for NeRF, where parameters directly represent a scene's 3D structure and are inherently sensitive. 
To enhance NeRF's utility while preserving privacy, a promising avenue is enabling the client side to efficiently fine-tune models after noise addition, representing a specific direction for research in privacy-preserving NeRF training.

\mypara{\securesplitnerf for Other NeRF Architectures}
In our paper, we adopt Instant-NGP~\cite{mueller2022instant}, a popular and efficient NeRF framework. 
Instant-NGP is widely used, as evidenced by state-of-the-art implementations like Zip-NeRF~\cite{barron2023zip}, which are based on its structure. 
Our proposed \splitnerf can also be applied to other MLP-based NeRF models, such as vanilla NeRF~\cite{mildenhall2020nerf} and D-NeRF~\cite{pumarola2021d}. 
\reconstructattack, \finetuneattack, and \securesplitnerf can be utilized with these similar \splitnerf architectures.

\mypara{Reconstruction Attacks against Split Learning}
Reconstruction attacks~\cite{zhu2019deep, fredrikson2015model, zhao2020idlg} typically necessitate access to full model parameters and gradients to reconstruct training data.
However, our attacks focus on recovering client model parameters and training data, which differs significantly. 
Although some studies~\cite{li2022auditing, zhang2023generative} have reduced the optimization space using GANs and over-parameterized networks, the optimization space for client model parameters remains extensive. In our paper, our attack methods leverage server model parameters and gradients effectively, demonstrating relative strength.


\mypara{The Practical Deployment Challenges of \securesplitnerf}
The main challenges of deploying \securesplitnerf come from two aspects, one is the frequent transmission of high-dimensional features and back-propagated gradients over bandwidth-limited wireless channels, and the other is the difficulty of training deep models on resource-constrained edge networks. Model pruning~\cite{lin2024split,deng2020model} directly eliminates the number of parameters of the neural network. Sparsification or quantization techniques can help compress intermediate embeddings and gradients, thereby reducing communication overhead~\cite{polino2018model, oh2023communication, zheng2023reducing}.  
For instance, the work~\cite{zheng2023reducing} achieves 34.96× compression with only a $2.9\%$ accuracy decrease on CIFAR-100 datasets. These solutions facilitate the deployment of \securesplitnerf.

\section{Related Work} \label{sec:relatedwork}

\mypara{Information Leakage in Split Learning}
Label leakage under split learning, initially investigated in the context of advertisement conversion prediction~\cite{li2021label}, involves an attack that leverages the differences in shared gradients between positive and negative data samples to achieve a high AUC score for inferring labels. 
However, this approach is restricted to binary classification.
The Unsplit attack~\cite{erdougan2022unsplit} employs a gradient-matching strategy to minimize the mean square error between the original gradients and those of a surrogate model to infer labels.
Additionally, Fu et al.~\cite{fu2022label} show that a malicious server can utilize a partially trained model to launch model completion attacks with few labeled data. 
The classic gradient matching~\cite{zhu2019deep} utilizes a learning-based method to infer the data and labels by minimizing the distance of gradients under the white-box setting (known model).
Similarly, the work~\cite{xie2023label} has a stronger attack setting to focus on the black-box regression problem setting, where the attacker does not know the target model.

Above all, all the previous works mainly utilize the difference of the gradients for the binary or multi-classification problem, or regression problem, which are not fit for our NeRF problem. 
To the best of our knowledge, we are the first to study the leakage against the split learning under the NeRF problem.

\mypara{Privacy protection in Split Learning}
To address these privacy concerns, protection is primarily achieved through two categories of methods. The first involves using advanced cryptographic protocols, such as \textit{secure multiparty computation}\cite{wagh2020falcon, mohassel2018aby3} and \textit{two-party computation}\cite{mohassel2017secureml, patra2021aby2, xie2021generalized}.
The second category employs perturbation-based methods that introduce randomness into the data-sharing process to protect privacy. 
This includes obfuscating the shared gradient information among parties or directly perturbing the information that needs protection~\cite{abadi2016deep, erlingsson2019amplification, ghazi2021deep, sun2023dpauc, yang2022differentially, xie2022differentially, qiu2023defending}. 
For example, Sun et al.~\cite{sun2023dpauc} focus on adding noise to gradients as a means to safeguard the underlying data from potential privacy breaches. 
Our proposed \securesplitnerf belongs to the perturbation defend methods, as we add noises to the gradients to protect privacy.

\section{Conclusion} \label{sec:conclusion}
In this paper, we propose the pioneering split-learning NeRF framework \splitnerf to address the NeRF training data privacy issues.
Concretely, we design an algorithm framework where the client and server can train NeRF collaboratively, while the client does not need to transfer local private scene data to the server.
While the \splitnerf framework appears inherently private, our sequentially devised \reconstructattack and \finetuneattack compromise the privacy of \splitnerf.
The experimental results demonstrate that the \splitnerf framework still has a serious privacy breach.
We further propose \securesplitnerf, secure \splitnerf, which can strike a trade-off between model utility and privacy protection in various scenarios.
Extensive experiments on three widely used public NeRF datasets illustrate the effectiveness of \securesplitnerf.

\begin{acks}
We would like to thank our shepherd and the anonymous reviewers for their insightful comments.
This work was supported in part by NSFC under grant 62336005; in part by the Shenzhen Science and Technology Program JCYJ20210324120011032; in part by the Guangdong Provincial Key Laboratory of Big Data Computing, the Chinese University of Hong Kong, Shenzhen.
\end{acks}

\bibliographystyle{ACM-Reference-Format}
\bibliography{bib}

\appendix

\begin{table*}[!htb] 
	\centering 
    \caption{\textbf{Attack experiment results of surrogate models with different structures. We use LPIPS-depth, LPIPS-gray, SSIM-depth, SSIM-gray, and Human-eval to measure attack effectiveness. $\uparrow(\downarrow)$ means a higher(lower) value is favored.
    }}
        \setlength{\tabcolsep}{0.5mm}
	\begin{tabular}{c c |c c c c c}  
        \toprule
        \textbf{Surrogate Model} & {\multirow{2}*{\textbf{Dataset}}} & \multicolumn{5}{c}{\textbf{Metric}} \\
	 \textbf{Structures} & ~ & LPIPS-depth $\downarrow$ & LPIPS-gray $\downarrow$ & SSIM-depth $\uparrow$ & SSIM-gray $\uparrow$ & Human-eval $\uparrow$ \\ 
        \midrule
        {\multirow{3}*{Two-layer color MLP}} & 3D-FRONT  & $0.26$ & $0.54$ & $0.87$ & $0.58$ & $5.0$\\
        ~ & Hypersim  & $0.40$ & $0.42$ & $0.83$ & $0.72$ & $5.0$\\
        ~ & ScanNet  & $0.20$ & $0.63$ & $0.92$ & $0.59$ & $4.6$\\
         \midrule
        {\multirow{3}*{Four-layer color MLP}} & 3D-FRONT  & $0.45$ & $0.54$ & $0.78$ & $0.66$ & $3.4$\\
        ~ & Hypersim & $0.48$ & $0.45$ & $0.79$ & $0.68$ & $5.0$\\
        ~ & ScanNet & $0.19$ & $0.65$ & $0.93$ & $0.58$ & $4.6$\\
        \bottomrule
	\end{tabular}
 \label{table:attack_different_surrogate}
\end{table*}

\begin{table*}[!htb] 
	\centering
    \vspace{0.2cm}
    \caption{\textbf{\securesplitnerf defense results against surrogate models with different structures.
    We use LPIPS-depth, LPIPS-gray, SSIM-depth, SSIM-gray, and Human-eval to measure \securesplitnerf's defense effectiveness. $\uparrow(\downarrow)$ means a higher(lower) value is favored.
    }}
    \setlength{\tabcolsep}{0.5mm}
	\begin{tabular}{c c |c c c c c}  
        \toprule
        \textbf{Surrogate Model} & {\multirow{2}*{\textbf{Dataset}}} & \multicolumn{5}{c}{\textbf{Metric}} \\
	 \textbf{Structures} & ~ & LPIPS-depth $\uparrow$ & LPIPS-gray $\uparrow$ & SSIM-depth $\downarrow$ & SSIM-gray $\downarrow$ & Human-eval $\downarrow$ \\ 
        \midrule
        {\multirow{3}*{Two-layer color MLP}} & 3D-FRONT  & $0.35$ & $0.47$ & $0.78$ & $0.71$ & $1.0$\\
        ~ & Hypersim  & $0.43$ & $0.59$ & $0.82$ & $0.80$ & $1.2$\\
        ~ & ScanNet  & $0.59$ & $0.68$ & $0.57$ & $0.56$ & $1.2$\\
         \midrule
        {\multirow{3}*{Four-layer color MLP}} & 3D-FRONT  & $0.71$ & $0.73$ & $0.53$ & $0.39$ & $1.0$\\
        ~ & Hypersim & $0.63$ & $0.69$ & $0.77$ & $0.67$ & $1.0$\\
        ~ & ScanNet & $0.56$ & $0.70$ & $0.56$ & $0.62$ & $1.2$\\
        \bottomrule
	\end{tabular}
\label{table:defense_different_surrogate}
\end{table*}

\section{Additional Experimental Results of Attack Ablation Study} 
\label{appendix:attack_ablation}

\begin{figure*}[!htb]
\centering
\includegraphics[width=180mm]{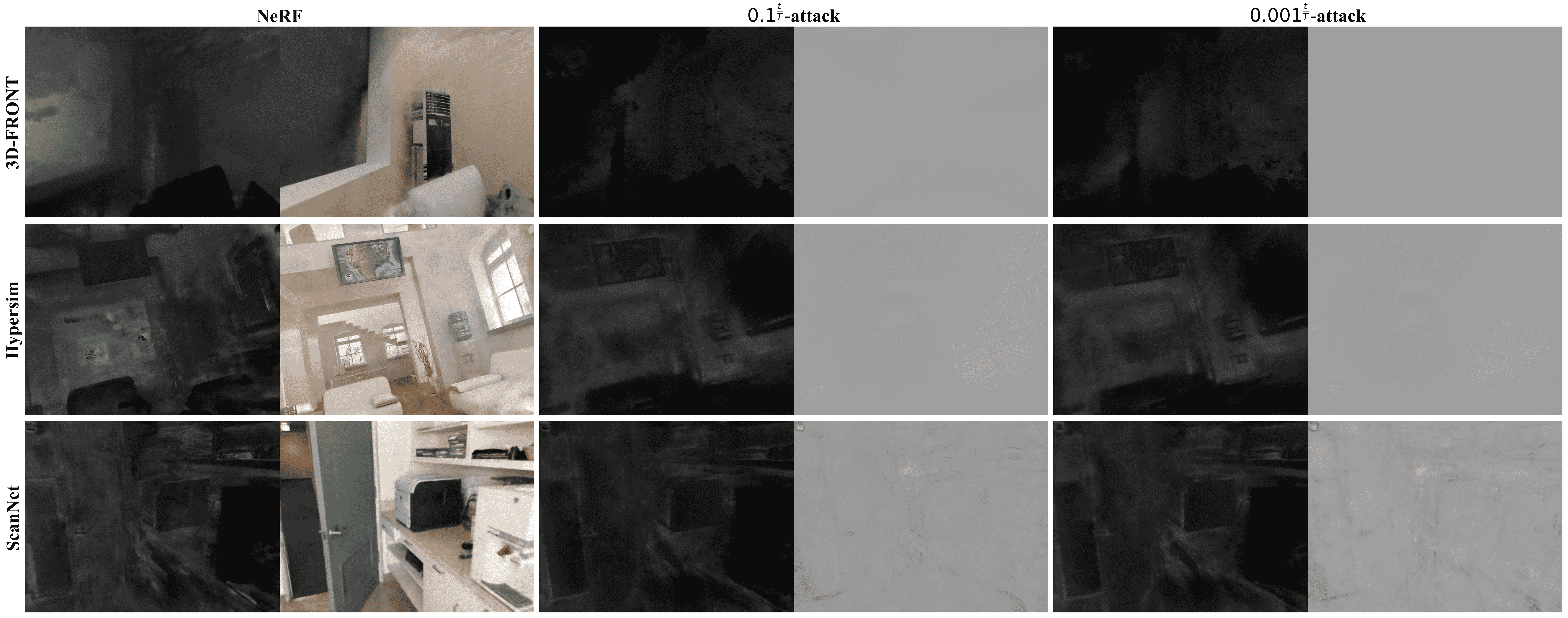}
\caption{\textbf{Comparison of \reconstructattack across the depth and color views on the three datasets, utilizing $0.1^\frac{t}{T}$ and $0.001^\frac{t}{T}$ learning rate decay schemes.
}}
\label{fig:lr_ablation_attackresult}
\end{figure*}

\begin{table*}[!htb]
	\centering \caption{\textbf{Defense method ablation study under $0.1^\frac{t}{T}$-attack.
    We evaluate different noise scales $c$ and decay ratio $r$.
    Experimental results indicate that \securesplitnerf with $c= 1.2, r = 0.0001$
    achieves a better trade-off between privacy and utility.
    We use PSNR and LPIPS to compare utility and use LPIPS-depth, LPIPS-gray, SSIM-depth, SSIM-gray, and Human-eval to measure privacy.
    We report the utility results without defense, using GT as the baseline.
    We highlight $c= 1.2, r = 0.0001$ method in the \colorbox{pink}{red} ground.
    $\uparrow(\downarrow)$ means a higher(lower) value is favored.
    }}
        \setlength{\tabcolsep}{0.5 mm}
	\begin{tabular}{c c | c c c c c c c c c c c c}  
        \toprule
        ~ & ~ & \multicolumn{12}{c}{\textbf{\securesplitnerf Configuration}} \\
        {\multirow{2}*{\textbf{Dataset}}} & {\multirow{2}*{\textbf{Metric}}} & $c = 0.6$ & $c =0.6$ & $c =0.6$ & $c = 1.2$ & $c = 1.2$ & $c = 1.2$ & $c = 2.4$ & $c = 2.4$ & $c = 2.4$ & $c = 4.8$ & $c = 4.8$ & {\multirow{2}*{\textbf{GT}}} \\ 
        ~ & ~ & $r = 0.0001$ & $r = 0.001$ & $r = 1$ & $r = 0.0001$ & $r = 0.001$ & $r = 1$ & $r = 0.0001$ & $r = 0.001$ & $r = 1$ &  $r = 0.0001$ & $r = 0.001$ &  \\
        \midrule
        {\multirow{7}*{3D-FRONT}} & LPIPS-depth $\uparrow$ & $0.46$ & $0.41$ & $0.60$ & $\cellcolor{pink}0.35$ & $0.49$ & $0.65$ & $0.46$ & $0.57$ & $0.71$ & $0.61$ & $0.58$ & - \\
        ~ & LPIPS-gray $\uparrow$ & $0.54$ & $0.54$ & $0.54$ & $\cellcolor{pink}0.54$ & $0.54$ & $0.54$ & $0.54$ & $0.54$ & $0.54$ & $0.54$ & $0.54$ & -\\ 
        ~ & SSIM-depth $\downarrow$ & $0.74$ & $0.81$ & $0.26$ & $\cellcolor{pink}0.77$ & $0.71$ & $0.41$ & $0.65$ & $0.63$ & $0.28$ & $0.46$ & $0.42$ & -\\ 
        ~ & SSIM-gray $\downarrow$ & $0.77$ & $0.77$ & $0.77$ & $\cellcolor{pink}0.77$ & $0.77$ & $0.77$ & $0.77$ & $0.77$ & $0.77$ & $0.77$ & $0.77$ & -\\ 
        ~ & Human-eval $\downarrow$ & $2.0$ & $1.6$ & $1.0$ & $\cellcolor{pink}1.4$ & $1.0$ & $1.0$ & $1.0$ & $1.0$ & $1.0$ & $1.0$ & $1.0$ & -\\
        ~ & PSNR $\uparrow$ & $21.06$ & $20.84$ & $18.88$ & $\cellcolor{pink}21.05$ & $20.75$ & $19.05$ & $20.82$ & $20.55$ & $17.08$ & $20.08$ & $19.13$ & $21.24$ \\
        ~ & LPIPS $\downarrow$ & $0.42$ & $0.47$ & $0.75$ & $\cellcolor{pink}0.43$ & $0.51$ & $0.79$ & $0.48$ & $0.60$ & $0.83$ & $0.69$ & $0.73$ & $0.39$ \\
        \midrule
        {\multirow{7}*{Hypersim}} & LPIPS-depth $\uparrow$ & $0.49$ & $0.51$ & $0.66$ & $\cellcolor{pink}0.52$ & $0.53$ & $0.64$ & $0.49$ & $0.61$ & $0.67$ & $0.53$ & $0.63$ & -\\
        ~ & LPIPS-gray $\uparrow$ & $0.65$ & $0.65$ & $0.65$ & $\cellcolor{pink}0.65$ & $0.65$ & $0.65$ & $0.65$ & $0.65$ & $0.65$ & $0.65$ & $0.65$ & -\\
        ~ & SSIM-depth $\downarrow$ & $0.69$ & $0.75$ & $0.54$ & $\cellcolor{pink}0.46$ & $0.74$ & $0.59$ & $0.73$ & $0.65$ & $0.37$ & $0.56$ & $0.51$ -\\ 
        ~ & SSIM-gray $\downarrow$ & $0.80$ & $0.80$ & $0.80$ & $\cellcolor{pink}0.80$ & $0.80$ & $0.80$ & $0.80$ & $0.80$ & $0.80$ & $0.80$ & $0.80$ & - \\ 
        ~ & Human-eval $\downarrow$ & $2.0$ & $1.0$ & $1.0$ & $\cellcolor{pink}1.6$ & $1.0$ & $1.0$ & $1.2$ & $1.0$ & $1.0$ & $1.0$ & $1.0$ & -\\  
        ~ & PSNR $\uparrow$ & $18.75$ & $18.66$ & $18.12$ & $\cellcolor{pink}18.72$ & $18.62$ & $17.81$ & $18.63$ & $18.57$ & $17.81$ & $18.49$ & $18.38$ & $18.77$ \\
        ~ & LPIPS $\downarrow$ & $0.43$ & $0.49$ & $0.77$ & $\cellcolor{pink}0.47$ & $0.58$ & $0.80$ & $0.56$ & $0.70$ & $0.80$ & $0.63$ & $0.75$ & $0.36$ \\
        \midrule
        {\multirow{7}*{ScanNet}} & LPIPS-depth $\uparrow$ & $0.52$ & $0.56$ & $0.60$ & $\cellcolor{pink}0.55$ & $0.55$ & $0.57$ & $0.51$ & $0.55$ & $0.57$ & $0.51$ & $0.57$ & -\\
        ~ & LPIPS-gray $\uparrow$ & $0.74$ & $0.75$ & $0.75$ & $\cellcolor{pink}0.75$ & $0.75$ & $0.75$ & $0.75$ & $0.75$ & $0.75$ & $0.75$ & $0.75$ & -\\
        ~ & SSIM-depth $\downarrow$ & $0.41$ & $0.35$ & $0.48$ & $\cellcolor{pink}0.38$ & $0.43$ & $0.36$ & $0.62$ & $0.59$ & $0.24$ & $0.66$ & $0.41$ & -\\ 
        ~ & SSIM-gray $\downarrow$ & $0.62$ & $0.62$ & $0.62$ & $\cellcolor{pink}0.62$ & $0.62$ & $0.62$ & $0.62$ & $0.62$ & $0.62$ & $0.62$ & $0.62$ & -\\ 
        ~ & Human-eval $\downarrow$ & $1.2$ & $1.0$ & $1.0$ & $\cellcolor{pink}1.0$ & $1.0$ & $1.0$ & $1.0$ & $1.0$ & $1.0$ & $1.0$ & $1.0$ & -\\ 
        ~ & PSNR $\uparrow$ & $19.66$ & $19.44$ & $17.94$ & $\cellcolor{pink}19.61$ & $19.39$ & $17.25$ & $19.51$ & $19.12$ & $17.35$ & $19.40$ & $18.65$ & $19.89$ \\
        ~ & LPIPS $\downarrow$ & $0.51$ & $0.58$ & $0.80$ & $\cellcolor{pink}0.55$ & $0.67$ & $0.80$ & $0.60$ & $0.72$ & $0.82$ & $0.67$ & $0.76$ & $0.42$ \\
        \bottomrule
	\end{tabular}
 \label{table:defense_ablation_0.1attack}
\end{table*}

\begin{table*}[!htb]
	\centering 
	\caption{\textbf{Defense method ablation study under $0.001^\frac{t}{T}$-attack.
    We evaluate different noise scales $c$ and decay ratio $r$.
    Experimental results indicate that \securesplitnerf with $c= 1.2, r = 0.0001$
    achieves a better trade-off between privacy and utility.
    We use PSNR and LPIPS to compare utility and use LPIPS-depth, LPIPS-gray, SSIM-depth, SSIM-gray, and Human-eval to measure privacy.
    We report the utility results without defense, using GT as the baseline.
    We highlight $c= 1.2, r = 0.0001$ method in the \colorbox{pink}{red} ground.
    $\uparrow(\downarrow)$ means a higher(lower) value is favored.
    }}
    \setlength{\tabcolsep}{0.50 mm}
	\begin{tabular}{c c | c c c c c c c c c c c c}  
        \toprule
        ~ & ~ & \multicolumn{12}{c}{\textbf{\securesplitnerf Configuration}} \\
        {\multirow{2}*{\textbf{Dataset}}} & {\multirow{2}*{\textbf{Metric}}} & $c = 0.6$ & $c =0.6$ & $c =0.6$ & $c = 1.2$ & $c = 1.2$ & $c = 1.2$ & $c = 2.4$ & $c = 2.4$ & $c = 2.4$ & $c = 4.8$ & $c = 4.8$ & {\multirow{2}*{\textbf{GT}}} \\ 
        ~ & ~ & $r = 0.0001$ & $r = 0.001$ & $r = 1$ & $r = 0.0001$ & $r = 0.001$ & $r = 1$ & $r = 0.0001$ & $r = 0.001$ & $r = 1$ &  $r = 0.0001$ & $r = 0.001$ &  \\
        \midrule
        {\multirow{7}*{3D-FRONT}} & LPIPS-depth $\uparrow$ & $0.34$ & $0.30$ & $0.59$ & $\cellcolor{pink}0.31$ & $0.41$ & $0.62$ & $0.38$ & $0.46$ & $0.57$ & $0.44$ & $0.51$ & -\\
        ~ & LPIPS-gray $\uparrow$ & $0.54$ & $0.54$ & $0.54$ & $\cellcolor{pink}0.54$ & $0.54$ & $0.54$ & $0.54$ & $0.54$ & $0.54$ & $0.54$ & $0.54$& -\\ 
        ~ & SSIM-depth $\downarrow$ & $0.82$ & $0.86$ & $0.65$ & $\cellcolor{pink}0.83$ & $0.76$ & $0.24$ & $0.74$ & $0.77$ & $0.57$ & $0.64$ & $0.62$ & -\\ 
        ~ & SSIM-gray $\downarrow$ & $0.77$ & $0.77$ & $0.77$ & $\cellcolor{pink}0.77$ & $0.77$ & $0.77$ & $0.77$ & $0.77$ & $0.77$ & $0.77$ & $0.77$ & -\\ 
        ~ & Human-eval $\downarrow$ & $3.4$ & $2.6$ & $1.0$ & $\cellcolor{pink}1.8$ & $1.0$ & $1.0$ & $1.6$ & $1.0$ & $1.0$ & $1.0$ & $1.0$ & -\\
        ~ & PSNR $\uparrow$ & $21.22$ & $20.92$ & $19.54$ & $\cellcolor{pink}21.14$ & $20.96$ & $17.77$ & $20.66$ & $20.21$ & $18.13$ & $20.15$ & $18.99$ & $21.24$ \\
        ~ & LPIPS $\downarrow$ & $0.41$ & $0.48$ & $0.72$ & $\cellcolor{pink}0.45$ & $0.51$ & $0.80$ & $0.49$ & $0.61$ & $0.79$ & $0.62$ & $0.72$ & $0.39$ \\
        \midrule
        {\multirow{7}*{Hypersim}} & LPIPS-depth $\uparrow$ & $0.38$ & $0.56$ & $0.68$ & $\cellcolor{pink}0.49$ & $0.53$ & $0.67$ & $0.50$ & $0.56$ & $0.69$ & $0.50$ & $0.63$ & -\\
        ~ & LPIPS-gray $\uparrow$ & $0.65$ & $0.65$ & $0.65$ & $\cellcolor{pink}0.65$ & $0.65$ & $0.65$ & $0.65$ & $0.65$ & $0.65$ & $0.65$ & $0.65$ & -\\
        ~ & SSIM-depth $\downarrow$ & $0.83$ & $0.69$ & $0.52$ & $\cellcolor{pink}0.76$ & $0.75$ & $0.49$ & $0.76$ & $0.66$ & $0.41$ & $0.74$ & $0.69$ & -\\ 
        ~ & SSIM-gray $\downarrow$ & $0.80$ & $0.80$ & $0.80$ & $\cellcolor{pink}0.80$ & $0.80$ & $0.80$ & $0.80$ & $0.80$ & $0.80$ & $0.80$ & $0.80$ & -\\ 
        ~ & Human-eval $\downarrow$ & $2.4$ & $2.2$ & $1.0$ & $\cellcolor{pink}1.6$ & $1.0$ & $1.0$ & $1.2$ & $1.0$ & $1.0$ & $1.0$ & $1.0$ & -\\  
        ~ & PSNR $\uparrow$ & $18.65$ & $18.66$ & $18.08$ & $\cellcolor{pink}18.70$ & $18.61$ & $17.83$ & $18.69$ & $18.51$ & $17.79$ & $18.65$ & $18.38$ & $18.77$ \\
        ~ & LPIPS $\downarrow$ & $0.42$ & $0.49$ & $0.76$ & $\cellcolor{pink}0.47$ & $0.58$ & $0.79$ & $0.52$ & $0.67$ & $0.80$ & $0.62$ & $0.75$ & $0.36$ \\
        \midrule
        {\multirow{7}*{ScanNet}} & LPIPS-depth $\uparrow$ & $0.49$ & $0.55$ & $0.57$ & $\cellcolor{pink}0.58$ & $0.54$ & $0.59$ & $0.50$ & $0.54$ & $0.63$ & $0.52$ & $0.53$ & -\\
        ~ & LPIPS-gray $\uparrow$ & $0.75$ & $0.75$ & $0.75$ & $\cellcolor{pink}0.75$ & $0.75$ & $0.75$ & $0.75$ & $0.75$ & $0.75$ & $0.75$ & $0.75$ & -\\
        ~ & SSIM-depth $\downarrow$ & $0.49$ & $0.33$ & $0.46$ & $\cellcolor{pink}0.29$ & $0.47$ & $0.21$ & $0.70$ & $0.57$ & $0.57$ & $0.56$ & $0.66$ & -\\ 
        ~ & SSIM-gray $\downarrow$ & $0.61$ & $0.61$ & $0.61$ & $\cellcolor{pink}0.61$ & $0.61$ & $0.61$ & $0.61$ & $0.61$ & $0.61$ & $0.61$ & $0.61$ & -\\ 
        ~ & Human-eval $\downarrow$ & $1.2$ & $1.2$ & $1.0$ & $\cellcolor{pink}1.0$ & $1.0$ & $1.0$ & $1.0$ & $1.0$ & $1.0$ & $1.0$ & $1.0$ & -\\ 
        ~ & PSNR $\uparrow$ & $19.73$ & $19.54$ & $18.04$ & $\cellcolor{pink}19.64$ & $19.22$ & $17.46$ & $19.51$ & $19.10$ & $17.03$ & $19.27$ & $18.78$ & $19.89$ \\
        ~ & LPIPS $\downarrow$ & $0.50$ & $0.57$ & $0.79$ & $\cellcolor{pink}0.55$ & $0.66$ & $0.80$ & $0.60$ & $0.73$ & $0.82$ & $0.68$ & $0.78$ & $0.42$ \\
        \bottomrule
	\end{tabular}
 \label{table:defense_ablation_0.001attack}
\end{table*}

\begin{figure*}[!htb]
\centering
\includegraphics[width=180mm]{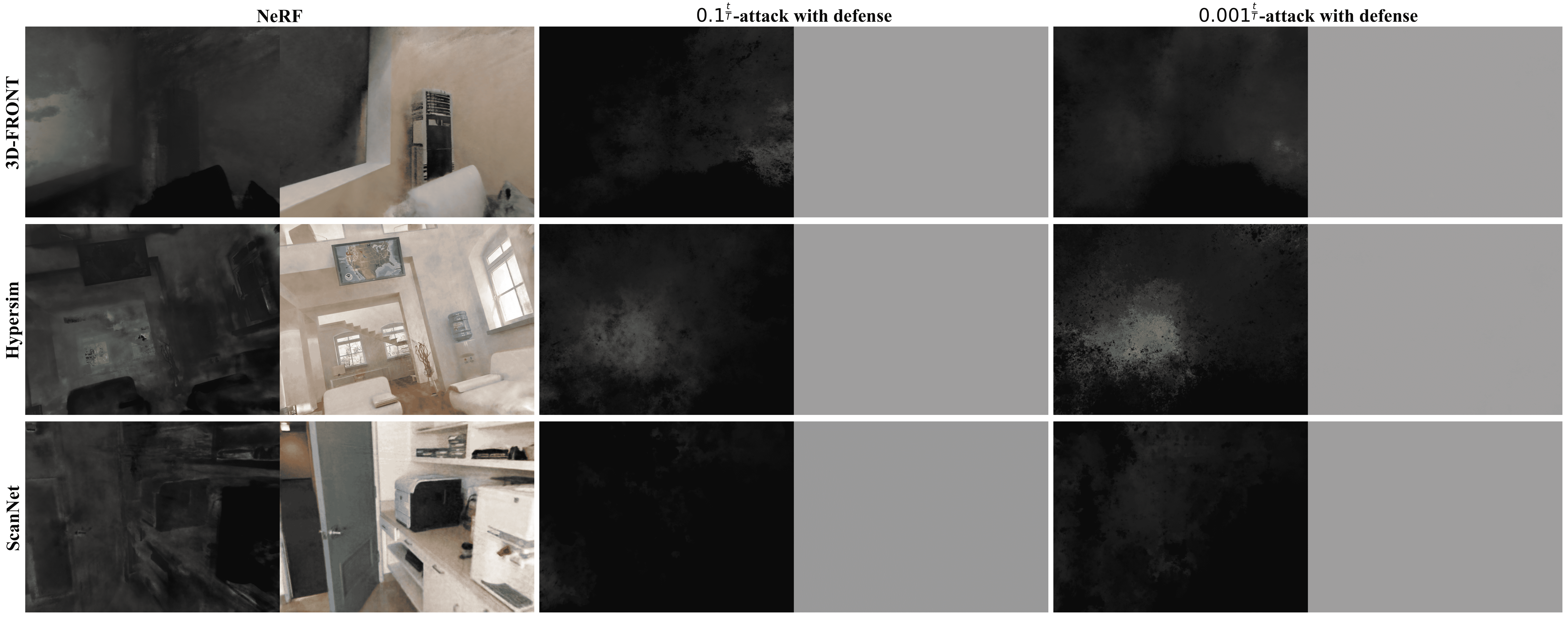}
\caption{\textbf{The results of \securesplitnerf under \reconstructattack utilizing $0.1^\frac{t}{T}$ and $0.001^\frac{t}{T}$ learning rate decay schemes.
}}
\label{fig:0.1_0.001_defense_result}
\end{figure*}

\begin{figure*}[!htb]
\centering
\includegraphics[width=180mm]{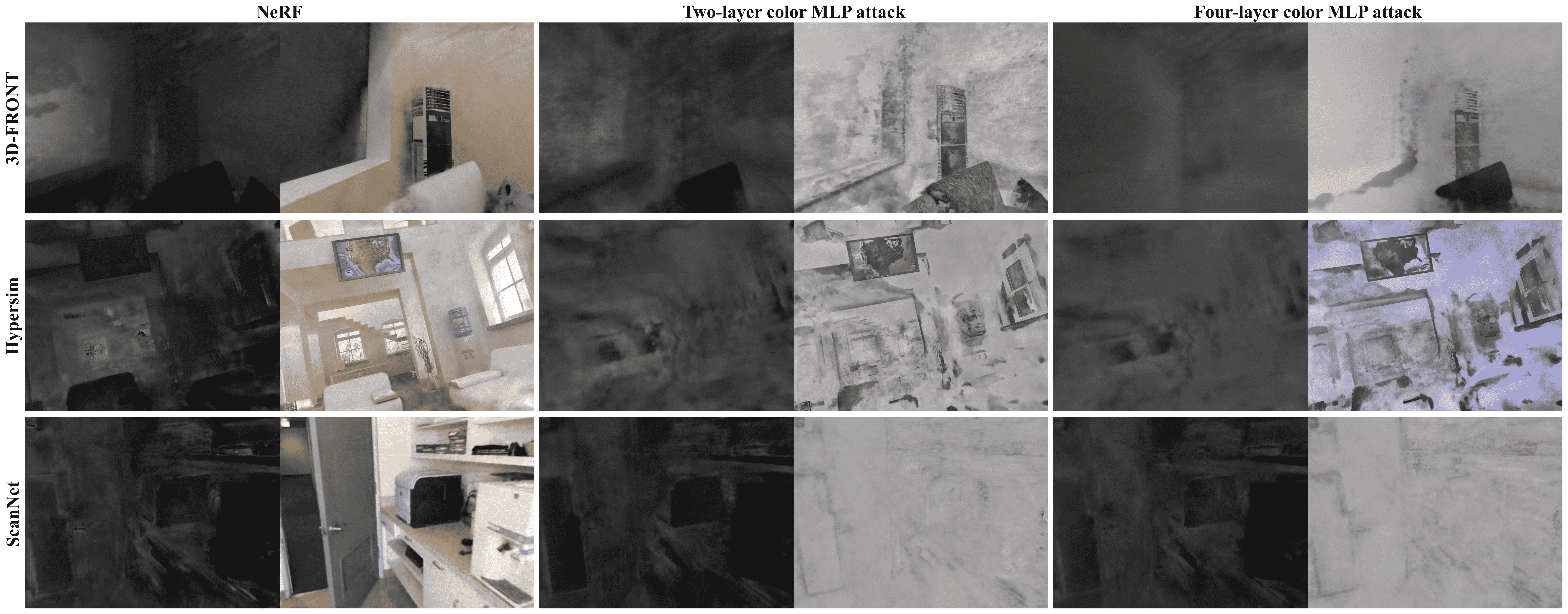}
\caption{\textbf{\reconstructattack results of different surrogate model structures: two-layer color MLP attack and four-layer color MLP attack.}}
\label{fig:attack_2and4colornet}
\end{figure*}

\begin{figure*}[!htb]
\centering
\includegraphics[width=180mm]{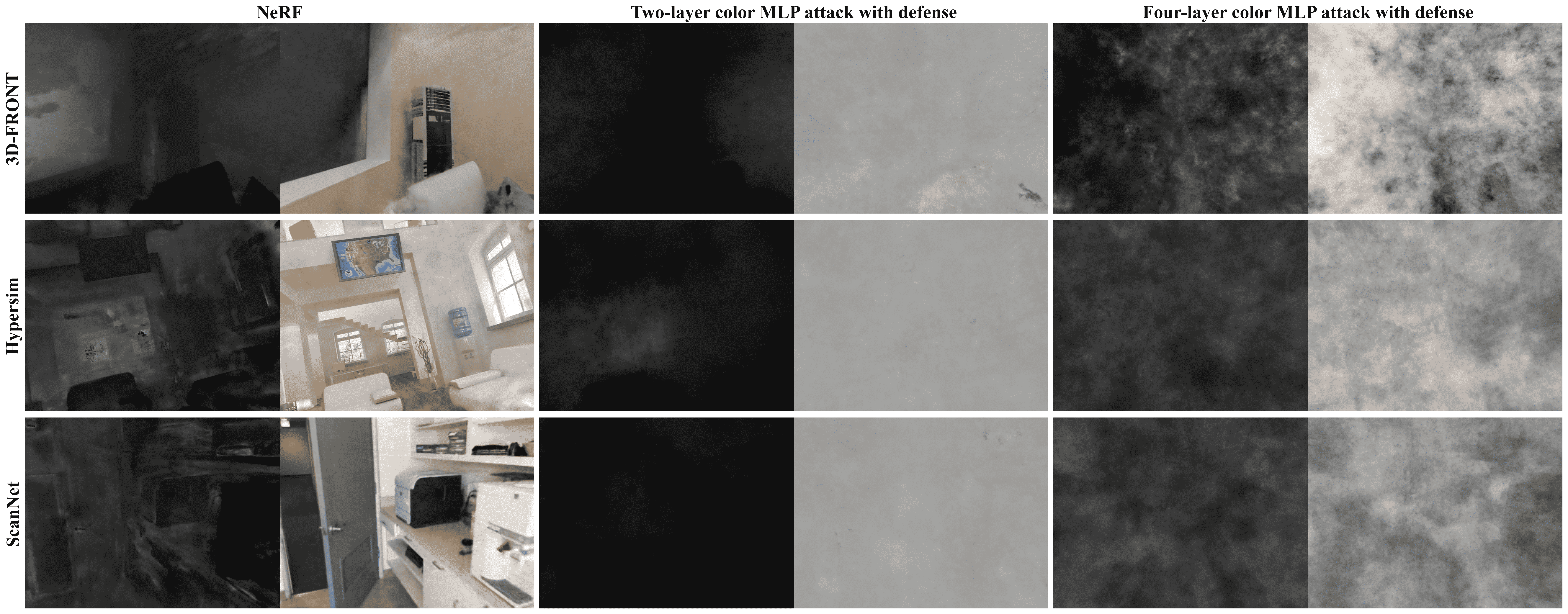}
\caption{\textbf{Defense results of \securesplitnerf with the configure $c=4.8, r= 0.001$ under two-layer color MLP attack and four-layer color MLP attack.
}}
\label{fig:defense_2and4colornet}
\end{figure*}

\begin{figure*}[!htb]
\centering
\includegraphics[width=180mm]{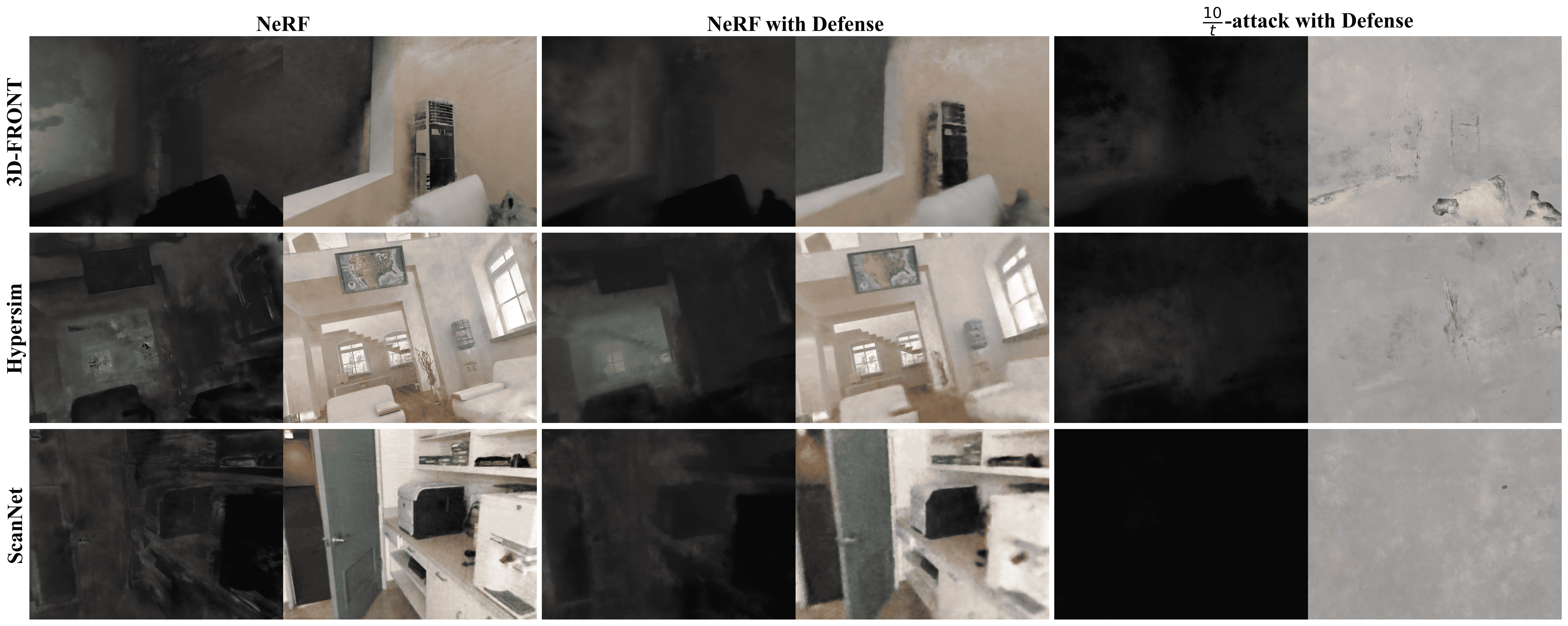}
\caption{\textbf{The light version \securesplitnerf Results. The light version \securesplitnerf maintains stable defense effectiveness and acceptable utility.}}
\label{fig:light_version_results}
\end{figure*}

\begin{figure*}[!htb]
\centering
\includegraphics[width=180mm]{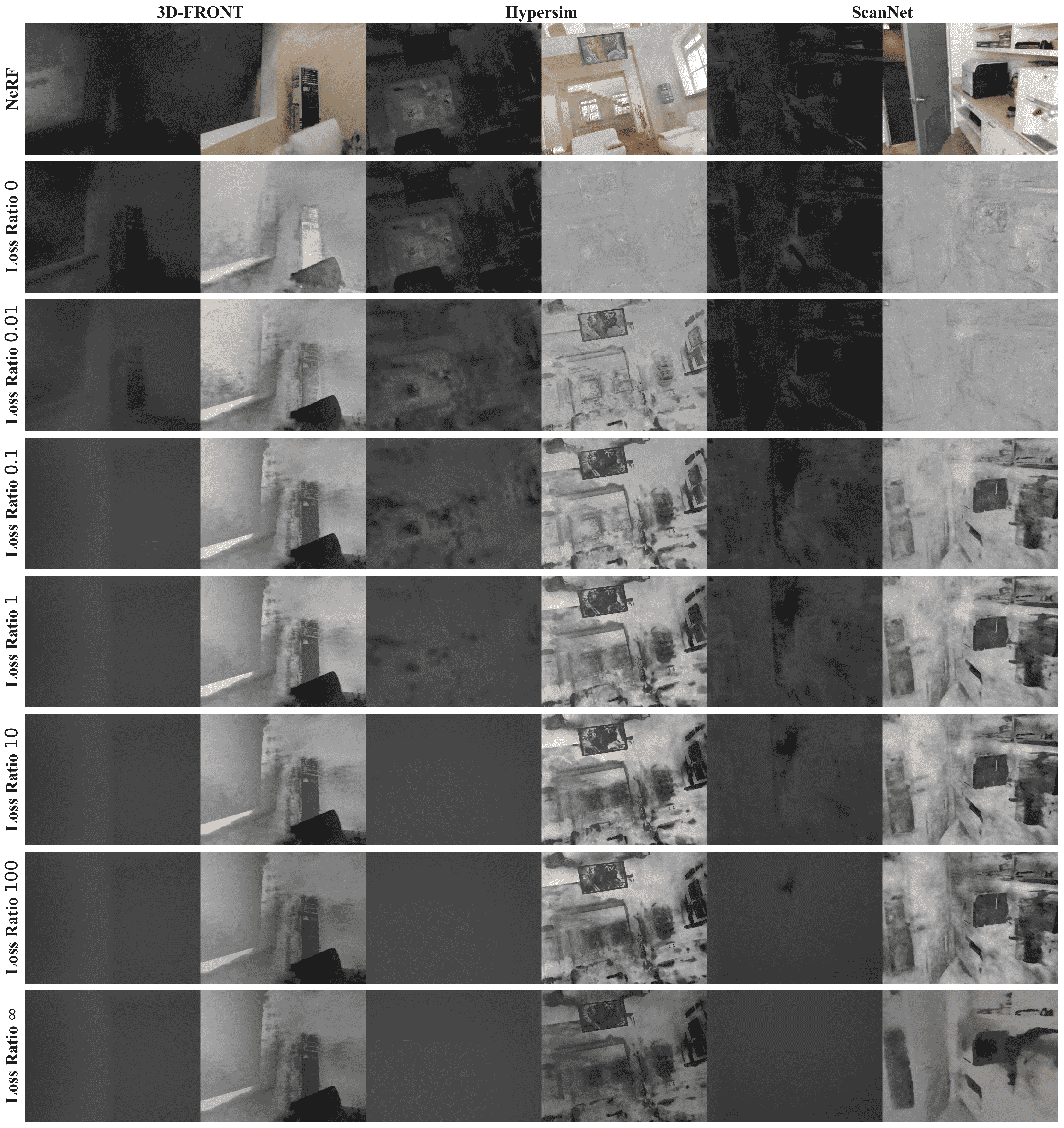}
\caption{\textbf{Comparison of \reconstructattack across the depth and color views on the three datasets, utilizing different loss ratios.
A loss ratio of $0$ indicates attacks solely on $L_{dummy}$, whereas a loss ratio of $\infty$ indicates attacks solely on $L_g$.
}}
\label{fig:lossratio_ablation_attackresult}
\end{figure*}

\mypara{Different Learning Rate Decay Schemes' Results}
\autoref{fig:lr_ablation_attackresult} presents the outcomes of the $0.1^{\frac{t}{T}}$-attack and $0.001^{\frac{t}{T}}$-attack, both employing a loss ratio of $0.01$. 
Generally, these attacks are slightly less effective compared to the $\frac{10}{t}$ attacks. 
They reconstruct some aspects of the original scene's depth perspective, such as parts of the sofa in the 3D-FRONT dataset, the water dispenser and picture frame in Hypersim, and the door outline in ScanNet. 
However, the level of detail and accuracy in these reconstructions falls short of what is achieved with the $\frac{10}{t}$-attack.

\mypara{Different Surrogate Model’s Loss Ratios' Results}
\autoref{fig:lossratio_ablation_attackresult} displays the performance of the $\frac{10}{t}$-attack across various loss ratios, ranging from $[0, 0.01, 0.1, 1, 10, 100, \infty]$. 
These attacks effectively capture scene information across different loss ratio settings. 
Notably, as the loss ratio increases from low to high, the attack's focus shifts from predominantly depth views to color views, illustrating a variation in attack impact influenced by the degree of model dependency on gradient information. 
Overall, under the $\frac{10}{t}$-attack, varying loss ratios consistently deliver potent attack outcomes, underscoring the effectiveness and robustness of this attack strategy.

\section{Additional Experimental Results of Surrogate Models with Different Structures}
\label{appendix:diff_surrogate_structures}

\mypara{Attack Experiments}
We conduct attack experiments with two additional surrogate model structures: a one-layer density MLP with a two-layer color MLP (referred to as the two-layer color MLP attack), and a one-layer density MLP with a four-layer color MLP (referred to as the four-layer color MLP attack). 
In each attack, we utilize $\frac{10}{t}$ learning rate decay schemes and the loss ratio of $0.01$. 
The results of these two attacks are shown in \autoref{table:attack_different_surrogate} and \autoref{fig:attack_2and4colornet}.

The results indicate that both attacks are highly effective. 
Specifically, the attacker's view successfully reveals the outline of the original scene. 
The human-eval metric is generally high, with most values exceeding $4.5$, indicating serious privacy disclosure.

\mypara{\securesplitnerf's Defense Effectiveness}
Based on the configuration of $c = 1.2, r = 0.0001$, we evaluate the defense effectiveness of \securesplitnerf against the two-layer color MLP attack and the four-layer color MLP attack. 
The defense results for these two attacks are shown in \autoref{table:defense_different_surrogate} and \autoref{fig:defense_2and4colornet}.

The results demonstrate that \securesplitnerf can effectively defend against attacks from different surrogate model structures. 
Visually, neither attack can recover any private information, with the attack views appearing nearly completely black or gray. 
From the perspective of attack metrics, the effectiveness of the attacks is significantly reduced, with the human-eval score dropping from about $4.5$ to around $1$.

\section{Additional Experimental Results of \securesplitnerf Ablation Study} 
\label{appendix:defense_ablation}

\mypara{\securesplitnerf's Defense Effectiveness against Other Attacks}
\autoref{fig:0.1_0.001_defense_result} demonstrates the defense effectiveness of \securesplitnerf against the $0.1^{\frac{t}{T}}$-attack and $0.001^{\frac{t}{T}}$-attack. 
The results clearly indicate that the attackers are unable to access any sensitive information about the scene, showcasing the robustness and universality of \securesplitnerf's defensive capabilities. 
This effectiveness is particularly pronounced when compared to the outcomes of successful attacks depicted in \autoref{fig:lr_ablation_attackresult}, highlighting the strength of \securesplitnerf's defense mechanisms.

\begin{figure*}[!htb]
\centering
\includegraphics[width=180mm]{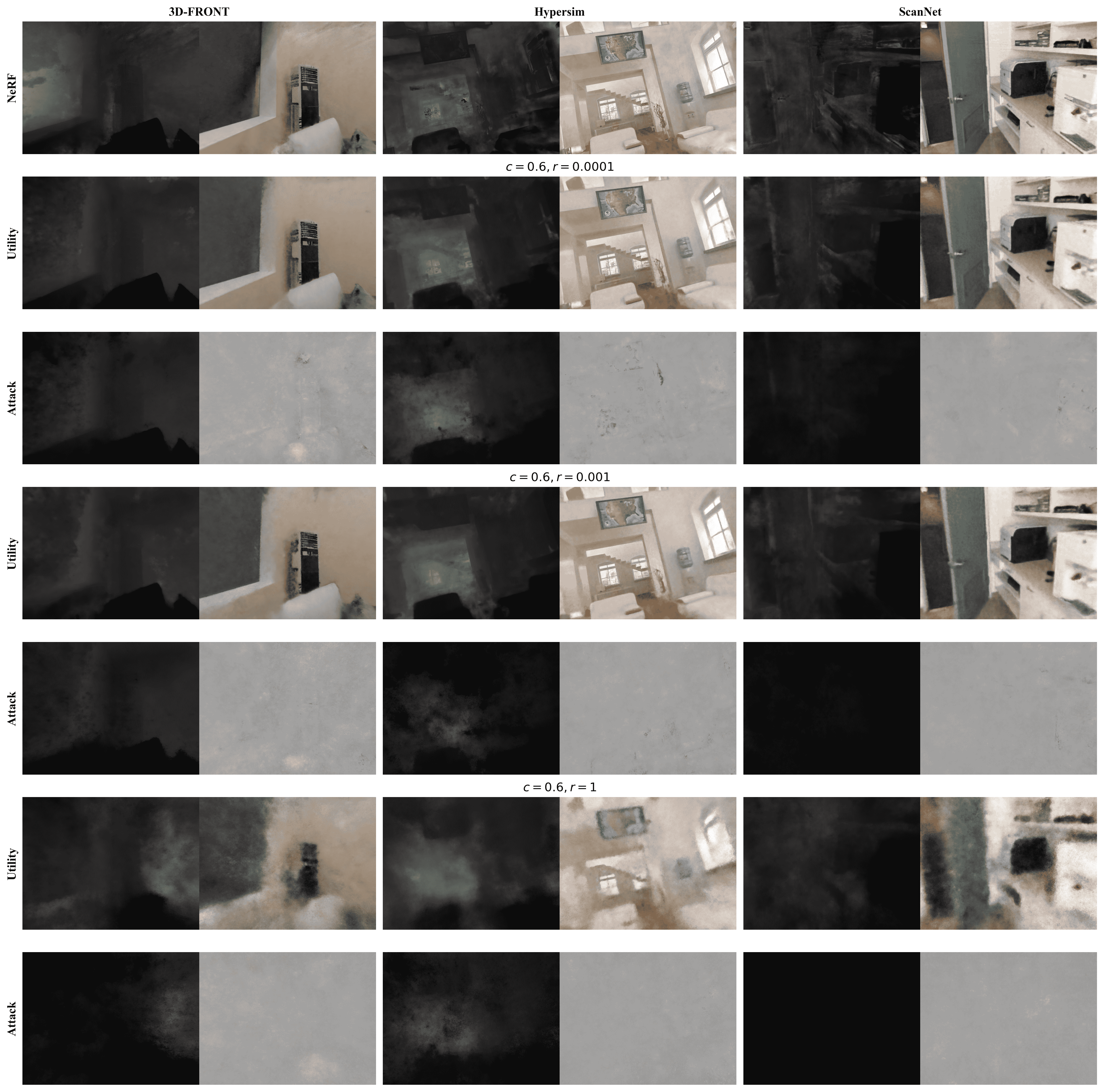}
\caption{\textbf{Experimental \securesplitnerf Results with the configuration $c = 0.6, r = 0.0001,0.001,1$.
}}
\label{fig:defense_noise_c_0.6_alldataset}
\end{figure*}

\begin{figure*}[!htb]
\centering
\includegraphics[width=180mm]{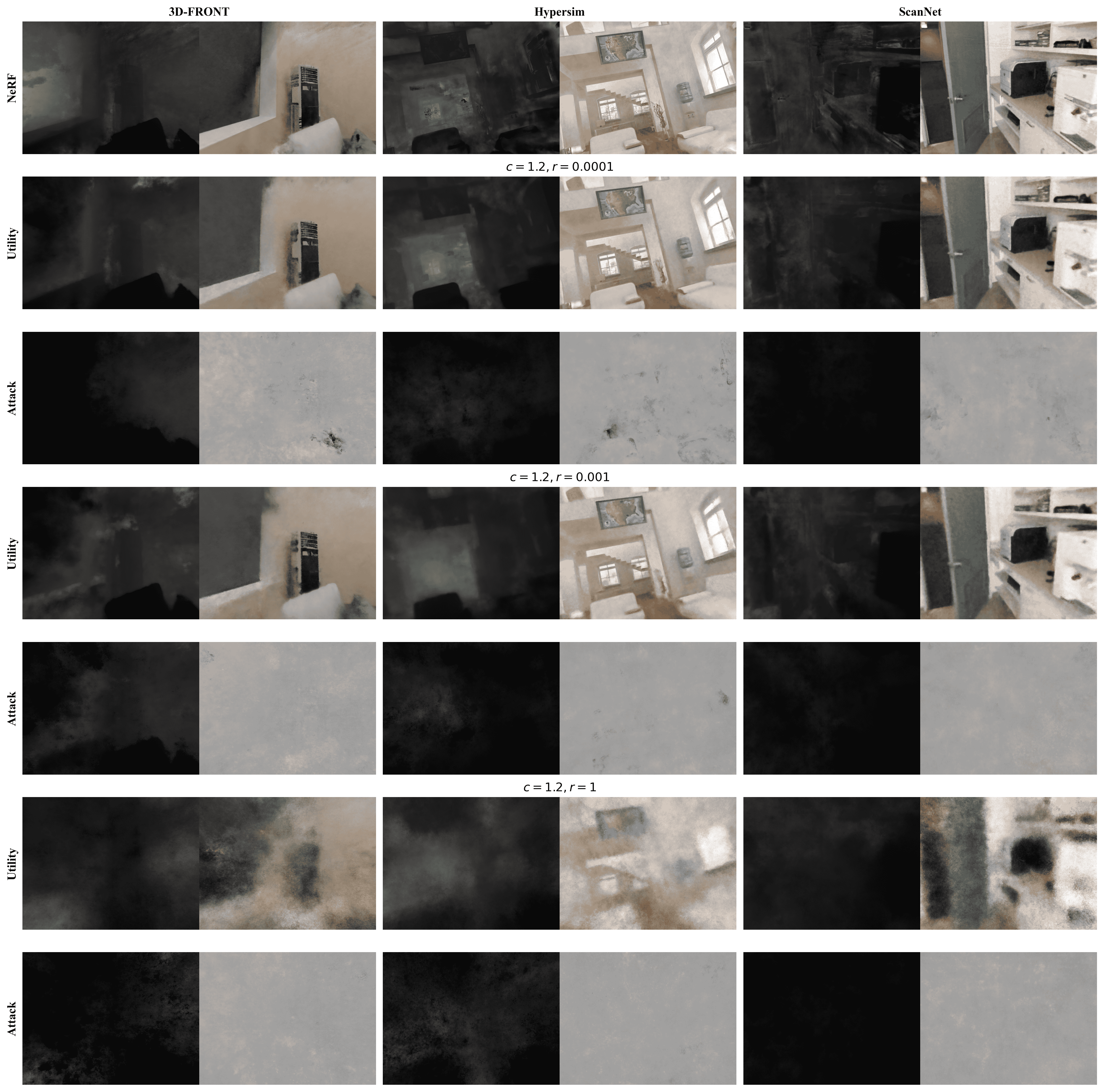}
\caption{\textbf{Experimental \securesplitnerf Results with the configuration $c = 1.2, r = 0.0001,0.001,1$.
}}
\label{fig:defense_noise_c_1.2_alldataset}
\end{figure*}

\begin{figure*}[!htb]
\centering
\includegraphics[width=180mm]{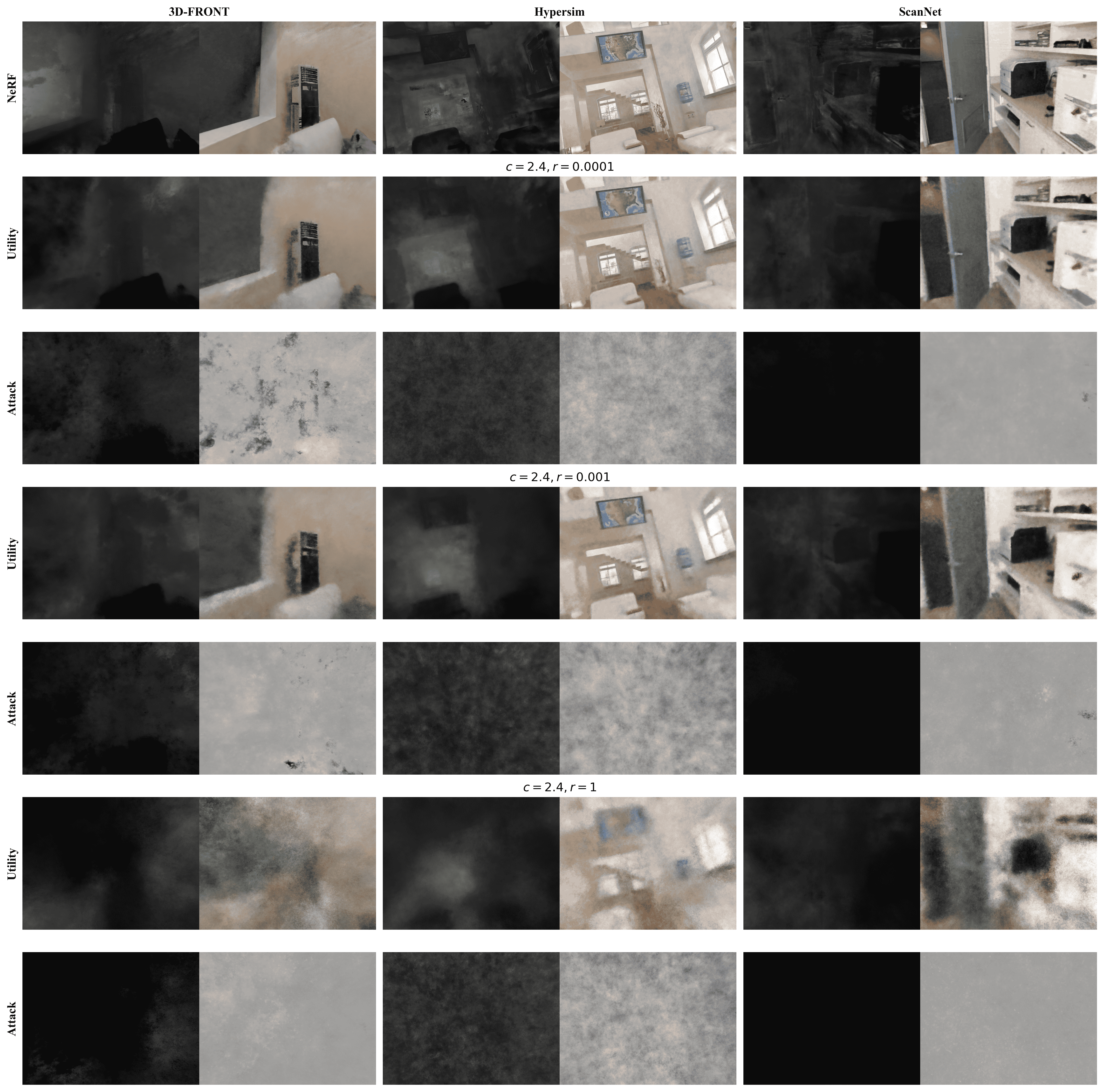}
\caption{\textbf{Experimental \securesplitnerf Results with the configuration $c = 2.4, r = 0.0001,0.001,1$.
}}
\label{fig:defense_noise_c_2.4_alldataset}
\end{figure*}

\begin{figure*}[!htb]
\centering
\includegraphics[width=180mm]{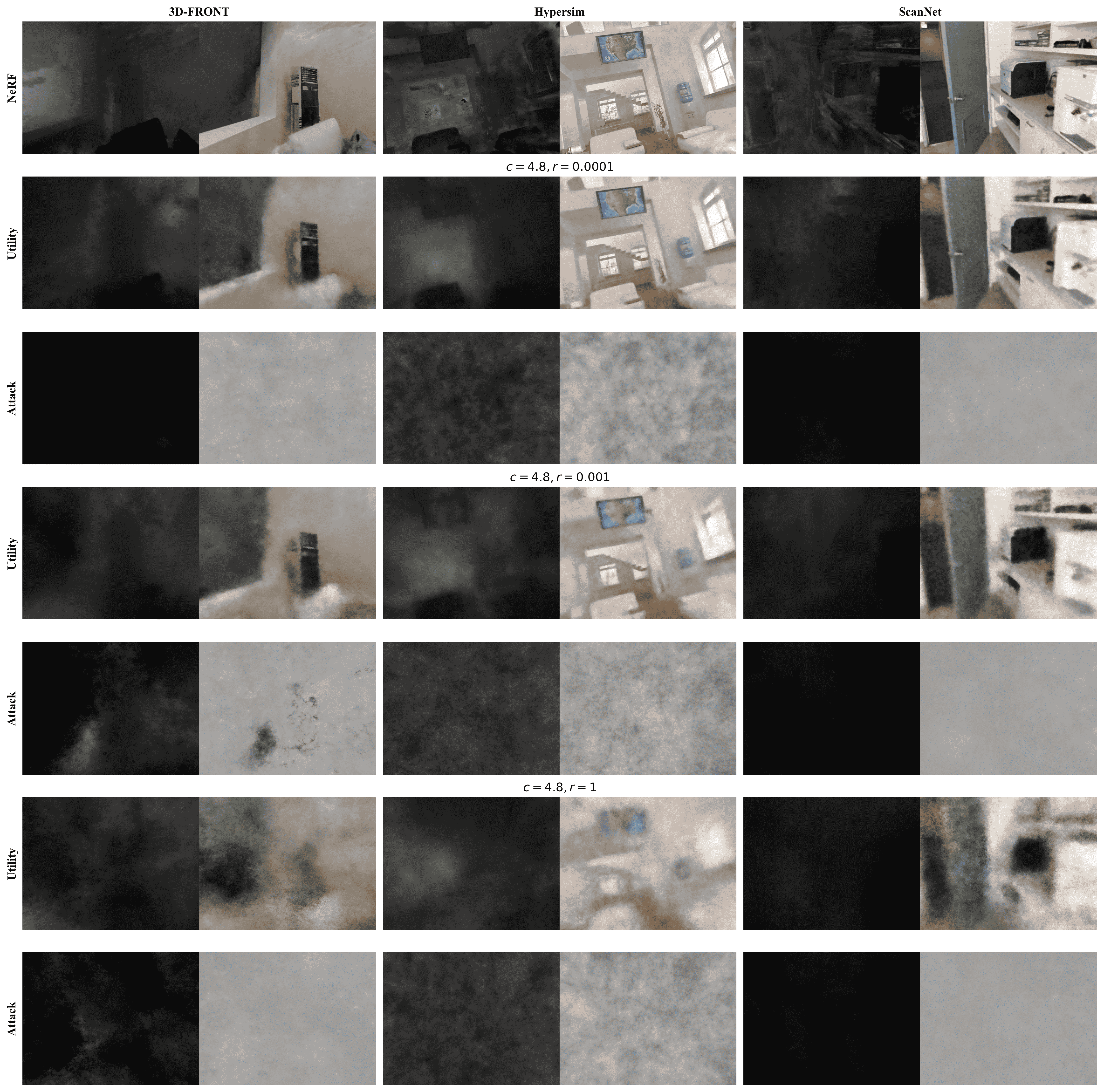}
\caption{\textbf{Experimental \securesplitnerf Results with the configuration $c = 4.8, r = 0.0001,0.001,1$.
}}
\label{fig:defense_noise_c_4.8_alldataset}
\end{figure*}

\mypara{Experimental Results of Different Configures of \securesplitnerf}
The performance of \securesplitnerf under various noise scales($c$) and decay ratios($r$) is documented across multiple figures and tables: \autoref{fig:defense_noise_c_0.6_alldataset}, \autoref{fig:defense_noise_c_1.2_alldataset}, \autoref{fig:defense_noise_c_2.4_alldataset}, \autoref{fig:defense_noise_c_4.8_alldataset}, \autoref{table:defense_ablation_0.1attack}, and \autoref{table:defense_ablation_0.001attack}. 

Overall, configurations with a decay rate of $0.0001$ exhibit better NeRF utility.
However, lower noise scale configs($c = 0.6$) are less effective in defending, allowing attackers to recover parts of the scene's outline still. 
For instance, in the 3D-FRONT dataset, as shown in \autoref{fig:defense_noise_c_0.6_alldataset} and \autoref{fig:defense_noise_c_1.2_alldataset}, attackers are able to discern outlines of sofas and air conditioners. 
On the other hand, a larger noise decay ratio $r$(i.e., $0.001, 1$) leads to slower noise decay, which, while resisting attacks, significantly diminishes model utility. 
The scene outlines become blurred, and the rendered images exhibit considerable noise.
Consequently, a configuration of \securesplitnerf with $c=1.2,r=0.0001$ strikes the optimal balance, effectively countering attacks while preserving good model utility. 

\section{Additional Experimental Results of \labelnoisenerf} 
\label{appendix:strawman_sol}

\begin{figure*}[!htb]
\centering
\includegraphics[width=180mm]{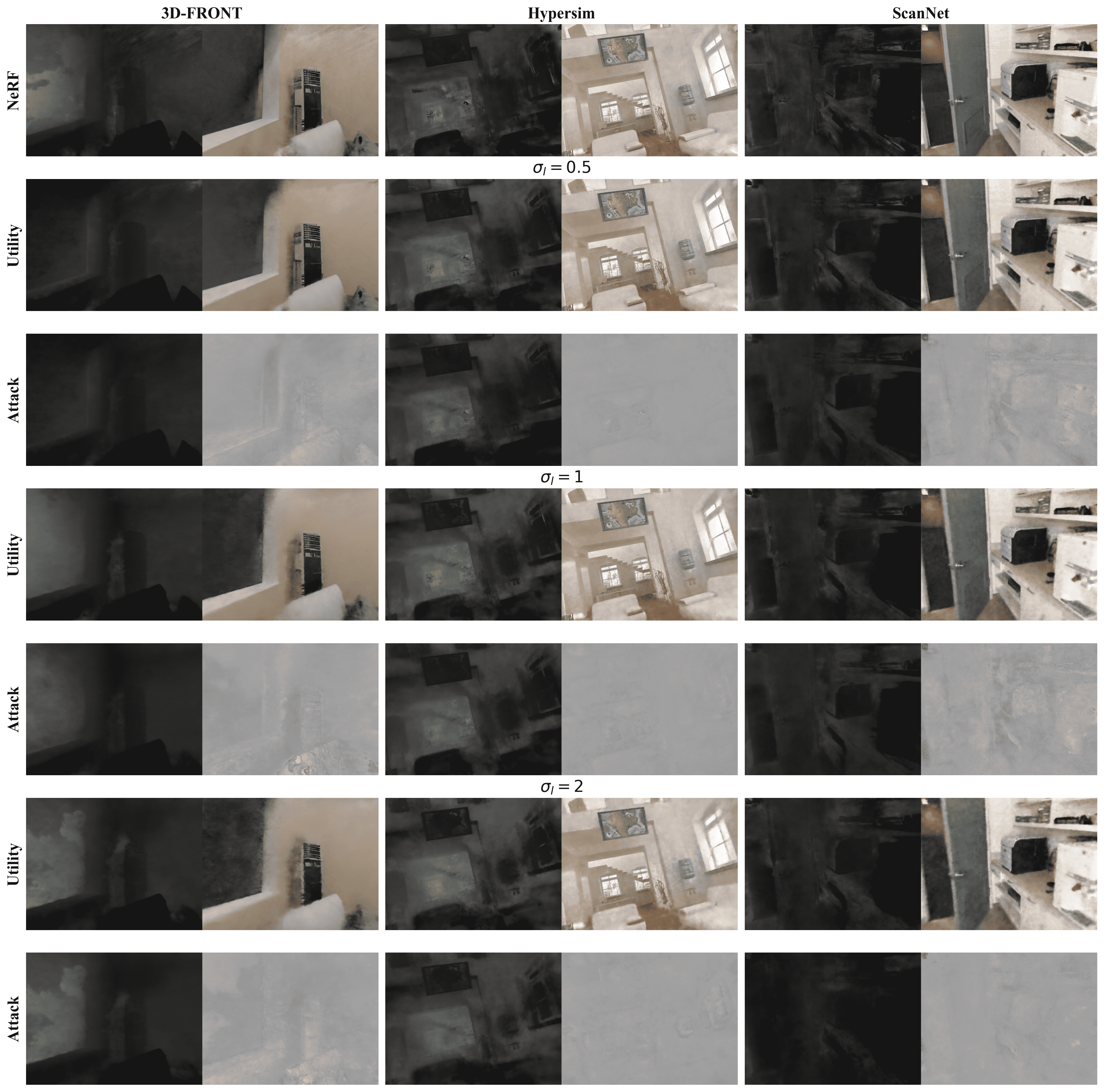}
\caption{\textbf{\labelnoisenerf Results with $\sigma_l = 0.5,1,2$.
}}
\label{fig:defense_noise_label_0.5_1_2_alldataset}
\end{figure*}

\begin{figure*}[!htb]
\centering
\includegraphics[width=180mm]{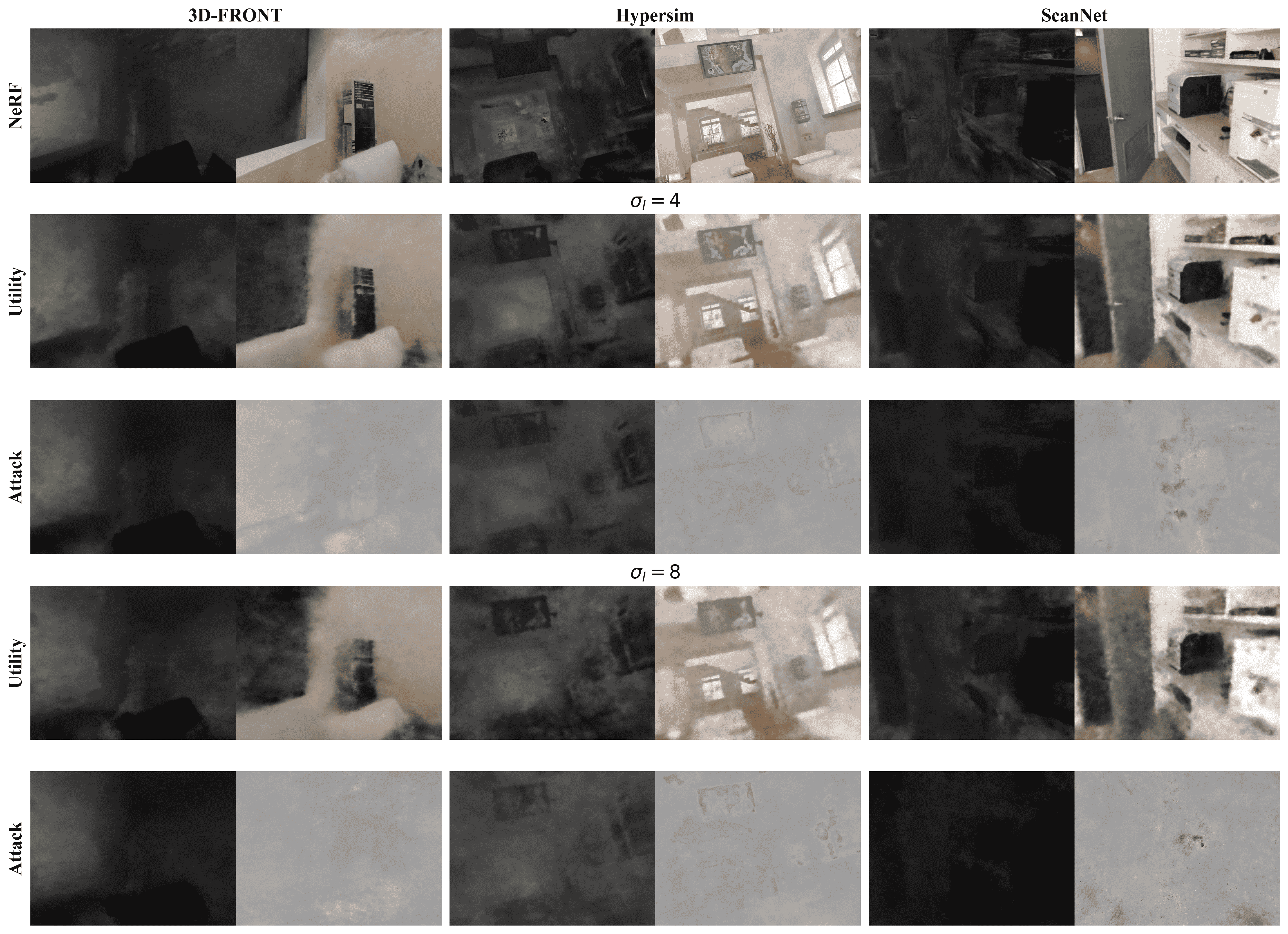}
\caption{\textbf{\labelnoisenerf Results with $\sigma_l = 4,8$.
}}
\label{fig:defense_noise_label_4_8_alldataset}
\end{figure*}

\autoref{fig:defense_noise_label_0.5_1_2_alldataset} and \autoref{fig:defense_noise_label_4_8_alldataset} display the performance of \labelnoisenerf under various noise scales $\sigma_l$.
The figures clearly demonstrate that \labelnoisenerf is ineffective. 
Even with high noise levels($\sigma_l=4,8$), the attacker is still able to successfully recover scene information. 
For example, under the Hypersim dataset, the outline of the picture frame and the room remain discernible despite the noise.

Additionally, the model's utility significantly deteriorates as the noise scale increases, leading to distorted scene colors that render the results unacceptable. 
This performance starkly contrasts with our proposed \securesplitnerf, underscoring its superior effectiveness and the inadequacies of \labelnoisenerf in providing robust defense while maintaining acceptable model utility.

\section{Experimental Results of Light Version \securesplitnerf}
\label{appendix:light_results}

The defense outcomes of the light version \securesplitnerf in \autoref{fig:light_version_results} are configured with $c = 1.2$ and $r = 0.0001$. 
Although the utility results for the light version are slightly worse than those of the standard version shown in \autoref{fig:defense-10-k}, they still maintain high availability. 
Additionally, the light version \securesplitnerf remains resistant to attacks.

\end{document}